\documentclass[11pt]{article}
\usepackage{amssymb,latexsym,amsmath}
\usepackage{xcolor}
\usepackage{bm}
\usepackage{amsmath,amscd}
\usepackage{graphicx}
\usepackage{newtxtext}
\usepackage{newtxmath}
\usepackage{geometry}
\usepackage{hyperref}
\hypersetup{
    colorlinks = true,
    urlcolor   = blue,
    citecolor  = black,
}
\newtheorem{theorem}{Theorem}[section]
\newtheorem{corollary}{Corollary}
\newcommand{\RomanNumeralCaps}[1]
\linenumbers

\title{Which symmetry group for elementary particles with an electric charge today and in the past?}

\author{\textbf{G. de Saxc\'e} \\
Univ. Lille, CNRS, Centrale Lille, UMR 9013 – LaMcube \\
Laboratoire de m\'ecanique multiphysique multi\'echelle, \\
F-59000, Lille, France, Email: gery.de-saxce@univ-lille.fr}

\begin{document}
\maketitle

\begin{abstract}
In this paper, we revisit the Kaluza-Klein theory from the perspective of the classification of elementary particles based on the coadjoint orbit method. The keystone conjecture is to consider the electric charge as an extra momentum on an equal footing with the mass and the linear momentum. We study the momentum map of the corresponding symmetry group $\hat{\mathbb{G}}_1$ which conserves the hyperbolic metric. We show that the electric charge is not an invariant, \textit{i.e.} it depends on the reference frame, which is in contradiction with the experimental observations. In other words, it is not the symmetry group of the Universe today as we know it. To avert this paradox, we scale the fifth coordinate and consider the limit when the cylinder radius $\omega$ vanishes. For the corresponding group $\hat{\mathbb{G}}_0$, the charge is an invariant then independent of the frame of reference and the observer. On this ground, we propose a cosmological scenario in which the elementary particles of the early Universe are classified from the momenta of the group $\hat{\mathbb{G}}_1$, next the three former dimensions inflate quickly while the fifth one shrinks, leading to the 4D era in which as today the particles are characterized by the momenta of the group $\hat{\mathbb{G}}_0$. By this mechanism, the elementary particles can acquire electric charge as a by-product of the $4 + 1$ symmetry breaking of the Universe. This work opens the way to the geometric quantization of charged elementary particles. 
We construct the corresponding $\hat{\mathbb{G}}_0$-connections by pullback on the space-time. Imposing that the 
linear 5-momentum is parallel-transported, we recover the conservation of the charge and the equation of motion with Lorentz force. We revisit the variational relativity and obtain the field equations for both the gravitation and  electromagnetic interactions with coupling terms which are negligible at the Newtonian approximation, allowing to recover Maxwell equations. 
\end{abstract}           


\vspace{0.2cm}






\section{Introduction}


Kaluza-Klein theory is an extension of general relativity to unify the gravitation and electromagnetism, considering a Riemannian manifold of dimension 5 by completing the 4 coordinates $(X^1, \cdots , X^4)$ of the space-time with an extra spacelike dimension coordinate $\hat{X}^5$.

Theodor Kaluza's pioneering paper, \cite{Kaluza 1915}, started out by interpreting the 1 + 5 metric as follows\footnote{Our convention in this paper is that convention that Greek indices run from 1 to 5 and Latin ones run from 1 to 4}
$$    \lbrack \hat{G}_{\mu\nu} \rbrack =  \left[ {{\begin{array}{cc}
       G_{ij} \hfill &  2 \, \alpha A_i  \hfill  \\
       2 \, \alpha A_j \hfill &  2 \, \phi \hfill  \\
   \end{array} }} \right]   
$$
where $G_{ij}$ is the space-time metric, $A_i$ is the electromagnetic potential, $\phi$ is a new field, often called dilaton, and $\alpha$ is a coupling constant. Besides, Theodor Kaluza introduce the so-called cylinder condition , {\it i.e.} the components of the metric and other tensors do not depend on the fifth coordinate. Even so, the expression of the contravariant metric $\hat{G}^{\alpha\beta}$ in terms of the  $\hat{G}_{\alpha\beta}$ being complex, he uses the linearized weak field approximation. The components of the Ricci tensor read
$$    \lbrack \hat{R}_{\mu\nu} \rbrack =  \left[ {{\begin{array}{cc}
       \partial_k \Gamma^k_{ij} - \partial_j \Gamma^k_{ik} \hfill &  - \alpha  \, \partial_k F_{ki}  \hfill  \\
       - \alpha \,  \partial_k F_{kj} \hfill &  - \square \, \phi \hfill  \\
   \end{array} }} \right]   
$$
where $F_{ij} = \partial_i A_j - \partial_j A_i$ is the electromagnetic field and $\square  = G^{ij} \nabla_i \nabla_j$. 
With the energy-momentum tensor of the dust matter $\hat{T}_{\mu\nu} = \rho_m \hat{U}_\mu \hat{U}_\nu$ as source, Einstein field equations extended to 5D
$$ \hat{R}_{ij} - \frac{1}{2} \, (\hat{G}^{kl} \hat{R}_{kl}) \, \hat{G}_{ij} 
  = - \kappa \, \hat{T}_{ij}
$$
allow to recover with $R_{ij}$ the field equations of the gravitation exactly,  Maxwell's equations also with $\hat{R}_{i5}$ but at the cost of an hypothesis that turned out, as remarked Einstein, to be untenable (grossly superluminal velocity of electrons in the fifth dimension). In contrast, Lorentz force appears naturally in the equation of motion of a particle but, because of the above hypothesis, the last term of this equation and also the equation with $\hat{R}_{55}$ are problematic.


After Kaluza's promising paper but suffering several shortcomings, another breakthrough is the work of Oskar Klein \cite{Klein 1926, Klein 1926 Nature} who adopts also the cylinder condition but with restrictions on the coordinate changes to $X^i = H^i (X'^1, \cdots, X'^4),\; \hat{X}^5 = \hat{X}'^5 + h (X'^1, \cdots, X'^4)$ and then the following form the metric
$$    \lbrack \hat{G}_{\mu\nu} \rbrack =  \left[ {{\begin{array}{cc}
       G_{ij} + \phi^2 A_i A_j \hfill &  \phi^2 A_i  \hfill  \\
       \phi^2 A_j \hfill &  \phi^2 \hfill  \\
   \end{array} }} \right]   
$$
which leads to a more tractable expression of the contravariant metric. In absence of source terms from the matter, the field equations are
$$ R_{ij} - \frac{1}{2} \, (G^{kl} R_{kl}) \, G_{ij}
   = \frac{1}{2} \, \kappa \, \phi^2 \lbrack 
     F^{ik} F^j_k - \frac{1}{4} \, (F^{kl} F_{kl}) \, G_{ij}
     \rbrack 
     - \frac{1}{\phi} \, \lbrack \nabla_i \partial_j \phi 
     - (\square \phi) \, G_{ij}
     \rbrack
$$  
$$ \nabla_k F^k_i + \frac{3}{\phi} \, (\partial_k \phi) F^k_i = 0
$$
$$ \square \phi = \frac{1}{2} \, \kappa^3 (F^{kl} F_{kl}) 
$$
One of the interests of the formulation is that the electromagnetic stress-energy tensor occurs naturally at the right hand member of the first equation. Another interest is that the second equation provides Maxwell equation when the dilaton $\phi$ is set to one, but if so the last equation is satisfied only if the right hand member is null, {\it i.e.} under the too restrictive constraint that the norms of the electric and magnetic fields are equal. In contrast, if the dilaton varies, the field equation has no classical interpretation. 
If we continue anyway to set $\phi$ to one, the equation of geodesics reads
$$ \nabla U^i = U_5 \, F^i_j U^j
$$
Owing to the cylinder condition, $U_5$ is constant along the geodesic, hence the idea to put $U_4 = q / m_0$ where $q$ is the electric charge and $m_0$ the rest mass, that leads to the equation of motion
$$ m_0 \nabla U^i = q \, F^i_j U^j
$$
where the right hand member is just Lorentz force.

However, the discoveries of the starting quantum mechanics suggested to Klein that the fifth dimension is curled up and microscopic. Using de Broglie wavelength $ \lambda^5 = h / p^5$ where $h$ is Planck constant and $p^5$ is the linear momentum in the fifth dimension, Klein found that the size $l_K$ of the space along the fifth dimension was about $10^{-31}$ cm, very small with respect to the nuclear dimension (of the order of $10^{-13}$ cm)
and thereby gave an explanation for the cylinder condition in this small value. For a comparison between Kaluza and Klein approaches, \cite{Sewards 2008} can be consulted. Classical theory was completed in the 1940s and the full field equations including the scalar field were obtained almost simultaneously (\cite{Thiry 1948a, Thiry 1948b, Jordan 1948, Scherrer 1949}).


Albert Einstein's position with respect to Kaluza-Klein theory has changed during his scientific career. In \cite{Einstein 1920}, when giving his inaugural lecture in Leyden, Einstein for the first time publicly commented positively on this unification program. However, in a \cite{Einstein 1922}, he regrets that   it cannot produce non-singular rotation symmetric particle solutions. Later on, his work with W. Mayer \cite{Einstein Mayer 1931} is a a first attempt to propose an unified field theory. Commenting their work,  \'E. Cartan \cite{Cartan 1934} interprets the torsion of the 5D space as the electromagnetic field. The last Einstein's works on the topics are in collaboration with P. Bergmann \cite{Einstein 1938} and later on with P. Bergmann and V. Bargmann \cite{Einstein 1941}. 
For more details, the reader is referred to \cite{van Dongen 2000}.

Despite of the astonishing breakthrough of the Kaluza-Klein theory (emergence of Lorentz force, Maxwell equations and electromagnetic energy stress tensor from the general relativity), often qualified of "Kaluza miracles", it suffers of several serious shortfalls not yet rectified. Nonetheless this theory has given rise to many developments that it is not possible to cite all. We are focused on the topics relevant to the issues dealt with in the present paper.



In the modern language of differential geometry, Oskar Klein's construction is formalized as a principal $U(1)$-bundle. The electromagnetic potential of covariant components $A_i$ can be interpreted as a connection 1-form valued in the Lie algebra of the group. From the viewpoint of the physics of elementary particles, a relevant generalization of the Kaluza-Klein theory consists in substituting  an arbitrary non-abelian group $G$ for $U(1)$. The connection 1-form has coordinates $A^a_j$ and the curvature form in a coordinate frame is
$$ F^a_{ij} = \frac{1}{2} \, (\partial_i A^a_j - \partial_j A^a_i)
             + C^a_{bc} A^b_i A^c_j
$$
where $C^a_{bc}$ are the structure constants of the group $G$. It must contain at least a $U(1)$ subgroup for the conservation of the electric charge and a $SU(2)$ subgroup for the conservation of the isospin \cite{Kerner 1981}.


Initially introduced by V. Bargmann in \cite{Bargmann} to solve cohomologic difficulties to construct group actions on quantum wave functions in the Galilean setting, the Lie group that is named after him acts on a 5D space of which the fifth coordinate has the physical meaning of a specific action (in the sense of Hamilton least action principle), then it has not to be confused with Kaluza-Klein space. Bargmannian geometric structures have been studies by several authors (\cite{Duval 1985, Duval 1991, de Saxce 2010, AffineMechBook, Cardall 2024}). Despite of the differences with Kaluza-Klein theory, some aspects of the Bargmannian approach were a source of inspiration for this work. 


Another issue which has grasped researcher attention is the Kaluza-Klein cosmology. A generalized Friedmann-Lema\^itre-Robertson-Walker metric in 5D were introduced in \cite{Chodos 1980}. The idea is that while the fifth dimension has been shrinking, the other three spatial dimensions have been expanding, that could explain the large ratio of the electromagnetic to gravitational forces as a consequence of the age of the Universe, in agreement with Dirac's large-number hypothesis. In \cite{Sahdev 1984}, Sahdev obtained improved solutions of the field equations with perfect fluids added on the right-hand side and found that this scenario can offer a resolution to the horizon problem. However, instead of stabilizing at some small but finite value, as any reasonable physics would require, the internal radius in the fifth dimension tended to zero. Other models attempting to stabilize asymptotically the internal radius were proposed in \cite{Matzner 1985, Copeland 1985, Okada 1986, Kerner 1988}. These interesting works did not necessarily receive the welcome they deserved, perhaps because they were based on the theory of Kaluza-Klein whose weaknesses were known. 

Finally, not developed from Kaluza-Klein theory but playing a key role in this work, we have to point out Jean-Marie Souriau's seminal  work in which he proposed 
for the classical description of an elementary particle to study a family of homogeneous spaces for a symmetry group, the coadjoint orbits, and to classify them (\cite{SSD, SSDEng}). On this ground, he introduced the geometric prequantization in \cite{Souriau 1966}. As Kaluza-Klein theory is based on an $U(1)$-principal bundle of which the base manifold is the space-time, the key idea of the prequantization is to work with a  $U(1)$-principal bundle of which the base space is a symplectic manifold, a coadjoint orbit of the structure group. A tentative to combine this method with the 5D approach to treat the mass and the electric charge on an equal footing is proposed in \cite{Sniatycki 1980} but, in contrast to Kaluza-Klein theory, the fifth dimension is timelike without physical justification. The reader not  familiar with this topic can consult, in addition to Souriau's references previously cited, Kostant \cite{Kostant 1970}  who is with Souriau one of the pioneers of this method, \cite{Kijowski 1977} and \cite{Woodhouse 1991}.

\vspace{1.2cm}


The paper is structured as follows.

Sections 2 and 3 cover the background material needed to tackle the core of the paper as well as to define the terminology and notations. The reader familiar with one or several of these topics can skim through them. \textbf{Section 2} is devoted to Euclidean spaces, pointing out adjoint maps, Hodge star operator and a generalization of the vector product to arbitrary dimension. In \textbf{Section 3}, we recall useful main results of symplectic mechanics, namely the coadjoint orbit method and the momentum map. The \textbf{Section 4} is a brief reminder of the application of this method to Poincar\'e group allowing to classify the elementary particles in relativity.

After these preliminaries, the main body of the paper is a trilogy: classification of the particles in Kaluza-Klein theory, pullback connection on the space-time, extended variational relativity. 

In \textbf{Section 5}, we apply the coadjoint orbit method 
to the symmetry group of Kaluza-Klein 5D space, denoted $\hat{\mathbb{G}}_1$, which conserves the hyperbolic metric. We show that the linear momentum along the fifth dimension, interpreted as the electric charge, is not preserved by the group, then depends on the reference frame, contradicting the observations. The aim of \textbf{Section 6} is to avert this paradox by using the fact that the fifth dimension is overwhelmingly small. The result of the zoom in is to reveal a new symmetry group denoted $\hat{\mathbb{G}}_0$ for which the charge is invariant. \textbf{Section 7} begins with the presentation of a cosmological scenario in which the particles in the early 5D Universe are classified according to the group $\hat{\mathbb{G}}_1$ and, after a transition phase in which the three former spatial dimensions inflate quickly while the fifth one shrinks, the particles in the 4D era are classified thanks to $\hat{\mathbb{G}}_0$. By this mechanism, the elementary particles can acquire electric charge as a by-product of the 4 + 1 symmetry breaking of the Universe. Next, we apply the coadjoint orbit method to the group  $\hat{\mathbb{G}}_\omega$ for the transition between $\hat{\mathbb{G}}_1$ and $\hat{\mathbb{G}}_0$. In 5D, we generalize the spin polarisation in the form of a plane. 

The second part of the trilogy is developed in  \textbf{Section 8}. Our aim is to construct $\hat{\mathbb{G}}_0$-connections on the frame principal bundle. As the zoom out leads to a singularity in the fifth dimension, we build on the space-time a pullback connection. 
We claim that the motion of a charged particle and the evolution of its charge are such that its linear 5-momentum is parallel-transported. The torsion free condition allows to show that the charge is conserved along the trajectory and we recover the Lorentz force.  

The trilogy ends with \textbf{Section 9} where we revisit Palatini variational relativity by adding to the ten potentials of the metric the four electromagnetic potentials. The generalized action depends on the matter and the connection representing both the gravitation and electromagnetic interactions. We discuss the field equations with coupling terms between the two interactions. At the Newtonian approximation, the coupling is weak and we recover Maxwell equations.

\section{Euclidean spaces}

The two former subsection are simple reminders to define the terminology and notations used in the paper. 

\subsection{Metric}

An Euclidean space $\mathcal{T}$ of dimension $n$ is a vector space equipped with a (covariant) metric $\bm{G}$, {\it i.e.} a nondegenerate 2-covariant tensor $\bm{G}$  
$$ \forall \bm{U}\in \mathcal{T},\qquad  
           \bm{G} (\bm{U}, \bm{V}) = 0 \quad \Leftrightarrow \quad
           \bm{V} = \bm{0}\ .
$$ 
The value of the metric tensor for $\bm{U}$ and $\bm{V}$ is called their scalar product and denoted $\bm{U}\cdot\bm{V}$. Let $(\bm{e}_i)$ be a basis of $\mathcal{T}$. The symmetric regular matrix $G$ gathering the components $G_{ij} = \bm{e}_i \cdot\bm{e}_j$ is called Gram's matrix. Thus, it holds:
$$ \bm{G} (\bm{U}, \bm{V}) = \bm{U}\cdot\bm{V} = U^T G\,V\ ,
$$ 
where $U$ and $V$ are the columns gathering respectively the components of $\bm{U}$ and $\bm{V}$. 
If $G$ is diagonal with element $+1$ or $-1$ on the diagonal, the basis is called orthonormal. The number $p$ of positive number on the diagonal is called the positive index of inertia. If $n  \geq 2$ and $p = 1$, the space is said hyperbolic. To every vector $\bm{U}$ is associated one and only one linear form $ \bm{V} \mapsto \bm{G} (\bm{U}, \bm{V})$ denoted $\bm{U}^*$. The covariant components of $\bm{U}^*$ depends on the contravariant components of $\bm{U}$ through the operation of lowering the index: $ U_i = G_{ij} U^j$. The elements $G^{ij}$ of the inverse $G^{-1}$ of Gram's matrix are the components of  a $2$-contravariant tensor $\bm{G}^{-1}$ called contravariant metric, hence the reverse operation of raising the index: $U^i = G^{ij} U_j$. 

\subsection{Adjoint of a linear map}

Let $\bm{A}$ be a linear map from an Euclidean space $\mathcal{T}_0$ into another one $\mathcal{T}$. Its adjoint (with respect to the scalar products) is the linear map $\bm{A}^*$ from $\mathcal{T}$ into $\mathcal{T}_0$ such that:
$$ \forall \bm{U}\in\mathcal{T},\quad \forall \bm{V}\in\mathcal{T}_0,\qquad 
    \bm{U} \cdot (\bm{A}\,\bm{V}) = (\bm{A}^*\bm{U}) \cdot  \bm{V}\ .
$$
If $\bm{A}$ is represented by the matrix $A$ in bases of $\mathcal{T}_0$ and $\mathcal{T}$, $\bm{A}^*$  is represented by:
\begin{equation}
   A^* = G_0^{-1} A^T G\ .
\label{A^* = G_0^(-1) A^T G} 
\end{equation}
We verify that:
$$ (\bm{A} + \bm{A}')^* = \bm{A}^* + \bm{A}'^*,\qquad 
    (\bm{A}\,\bm{B})^* = \bm{B}^* \bm{A}^*,\qquad 
    (\bm{A}^*)^* = \bm{A}\ .
$$
In particular, if $\mathcal{T}_0 = \mathbb{R}$, the linear map $\bm{U}: \mathbb{R} \rightarrow \mathcal{T}: \lambda \mapsto \lambda\,\bm{U}$ can be identified to the vector $\bm{U}$ and $\bm{U}^*$ is the unique linear form associated to $\bm{U}$ with respect to the metric since (\ref{A^* = G_0^(-1) A^T G}) degenerates into
\begin{equation}
     U^* = U^T G
\label{U^* = U^T G}
\end{equation}
which is the matrix form of index lowering. Then the scalar product of two vectors reads: $\bm{U}\cdot\bm{V} = U^* \,V$. Another particular case of interest is when $\mathcal{T}_0 = \mathcal{T}$ then (\ref{A^* = G_0^(-1) A^T G}) reads:
\begin{equation}
   A^* = G^{-1} A^T G
\label{A^* = G^(-1) A^T G} 
\end{equation}
We verify that $Tr (\bm{A}^*) = Tr (\bm{A})$. The linear map is self-adjoint (resp. anti-self-adjoint or skew-adjoint) if:
$$ \bm{A} = \bm{A}^*\qquad (\mbox{resp.}\quad \bm{A} = - \bm{A}^*)\ .
$$

\subsection{Vector product}

We consider an oriented Euclidean space $\mathcal{T}$ of dimension $n$, {\it i.e.} there exists a volume form $vol \in \bigwedge^n \mathcal{T}^*$ such that $vol(\bm{e}_1, \ldots, \bm{e}_{n}) = 1$
for every orthonormal basis $(\bm{e}_i)$. We denote  $vol(\bm{e}_1, \ldots, \bm{e}_{q})$  the $(n - q)$-form such that 
$$ (vol(\bm{e}_1, \ldots, \bm{e}_{q})) (\bm{e}_{q + 1}, \ldots, \bm{e}_{n}) = vol (\bm{e}_1, \ldots, \bm{e}_{q},\bm{e}_{q + 1}, \ldots, \bm{e}_{n})
$$
We call \textbf{vector product} of $(n - 1)$ vectors $\bm{V}_1, \ldots, \bm{V}_{n - 1}$ the vector $\mathcal{J}(\bm{V}_1, \ldots, \bm{V}_{n - 1})$ such that 
$$ \mathcal{J}
    (\bm{V}_1, \ldots, \bm{V}_{n - 1})^* \bm{U} 
    = vol(\bm{V}_1, \ldots, \bm{V}_{n - 1},\bm{U}) 
$$
In \S 26.B of \cite{CL}, it is proved that

\begin{theorem}[Souriau]
Properties of the vector product:
\begin{itemize}
\item [$\diamondsuit$] $\mathcal{J}(\bm{V}_1, \ldots, \bm{V}_{n - 1}) \neq \bm{0}$ if and only if  $\bm{V}_1, \ldots, \bm{V}_{n - 1}$  are linearly independent
\item [$\heartsuit$] $\mathcal{J}(\bm{V}_1, \ldots, \bm{V}_{n - 1})$ is orthogonal to every argument $\bm{V}_i$ of $\mathcal{J}$
\item [$\spadesuit$] The linear map $\mathcal{J}(\bm{V}_1, \ldots, \bm{V}_{n - 2}) : \mathcal{T} \rightarrow \mathcal{T} : \bm{V} \mapsto \mathcal{J}(\bm{V}_1, \ldots, \bm{V}_{n - 2}, \bm{V})$ is one-to-one and skew-adjoint
\end{itemize}
\label{thm vector product}
\end{theorem}

Denoting $vol_{i_1 \ldots i_n}$ the covariant components of  the volume form of an oriented Euclidean space of dimension $n$ and $V^i_r$ the $i$-th component of the vector $\bm{V}_r$, the $j$-th component of the vector product is given by
$$ (\mathcal{J} (\bm{V}_1, \ldots, \bm{V}_{n - 1}))^j 
   = \sum_{i_1 \ldots i_{n - 1}} \, vol_{i_1 \ldots i_{n - 1}}\, ^j \,\, V^{i_1}_1 \ldots V^{i_{n - 1}}_{n - 1}
$$
where we rise the last index of the volume form with the metric.
The element of the matrix $\mathcal{J} (\bm{V}_1, \ldots, \bm{V}_{n - 2})$ at the intersection of the $j$-th row and the $k$-th column  is
$$ (\mathcal{J} (\bm{V}_1, \ldots, \bm{V}_{n - 2}))^j_k 
 =  \sum_{i_1 \ldots i_{n - 1}} \, vol_{i_1 \ldots i_{n - 2} k}\, ^j \,\, V^{i_1}_1 \ldots V^{i_{n - 2}}_{n - 2}
$$

Now, we would like to obtain a recursive formula to calculate the vector product when we add an extra dimension.
If $V, V_i \in \mathcal{T} = \mathbb{R}^n$ , we put
$$ \hat{V} = \left[ {{\begin{array}{cc}
       V \hfill  \\
       v  \hfill  \\
   \end{array} }} \right] ,
   \hat{V}_i = \left[ {{\begin{array}{cc}
       V_i \hfill  \\
       v_i  \hfill  \\
   \end{array} }} \right] \in  \hat{\mathcal{T}} = \mathbb{R}^{n + 1}
$$
Then, it can be proved that
\begin{equation}
     \hat{\mathcal{J}}(\hat{V}_1, \ldots, \hat{V}_n) = \left[ \sum^n_{k=1}
(-1)^{n - k + 1}\, v_k  \, \mathcal{J}(V_1, \ldots, V_{k - 1}, V_{k + 1}, \ldots, V_n)^*,
vol (V_1, \ldots, V_n) \right]^*
\label{hat(J) =}
\end{equation}

In the particular case of the classical positive Euclidean space of dimension 3 ($n = p = 3$), we use the notation $j$ instead of $\mathcal{J}$ and we recover the cross product $j(u,v) = u \times v$ and the linear map $u \mapsto j(u)$ is one-to-one from $\mathbb{R}^3$ into the space of $3 \times 3$ skew-symmetric matrices. In the sequel, the notation $\mathcal{J}$ will be reserved to the dimension 4 and $\hat{\mathcal{J}}$ to the dimension 5. In $\mathbb{R}^4$ equipped with the metric of Gram's matrix $G = diag(1, -1, -1, -1)$, the vector product of 3 vectors 
$$ \Pi_k = \left[ {{\begin{array}{cc}
       m_k \hfill  \\
       p_k  \hfill  \\
   \end{array} }} \right], \qquad m_k \in \mathbb{R}, \quad p_k \in \mathbb{R}^3 \qquad (1 \leq k \leq 3)
$$ 
itemizes as
\begin{equation}
   \mathcal{J}(\Pi_1, \Pi_2, \Pi_3) =  \left[ {{\begin{array}{cc}
       vol (p_1, p_2, p_3) \hfill  \\
        m_1 j (p_2, p_3) - m_2 j (p_1, p_3) + m_3 j (p_1, p_2) \hfill  \\
   \end{array} }} \right]   
\label{J (Pi_1, Pi_2, Pi_3) =}
\end{equation}
from which we deduce 
\begin{equation}
   \mathcal{J}(\Pi_1, \Pi_2) =  \left[ {{\begin{array}{cc}
       0            \hfill &  j(p_1, p_2)^T                   \hfill  \\
        j(p_1, p_2) \hfill &  m_1 j (p_2) - m_2 j (p_1) \hfill  \\
   \end{array} }} \right]   
\label{J (Pi_1, Pi_2) =}
\end{equation}
In $\mathbb{R}^5$ equipped with the metric of Gram's matrix $\hat{G} = diag (1, -1, -1, -1, - \omega^2 )$, let us consider 3 vectors 
$$ \hat{\Pi}_k = \left[ {{\begin{array}{cc}
       \Pi_k \hfill  \\
       q_k  \hfill  \\
   \end{array} }} \right], \qquad \Pi_k \in \mathbb{R}^4, \quad  q_k \in \mathbb{R} \qquad (1 \leq k \leq 3)
$$ 
Then
\begin{equation}
   \hat{\mathcal{J}}(\hat{\Pi}_1, \hat{\Pi}_2, \hat{\Pi}_3) =  \left[ {{\begin{array}{cc}
        q_1 \mathcal{J} (\Pi_2, \Pi_3)  - q_2 \mathcal{J} (\Pi_1, \Pi_3) + q_3 \mathcal{J} (\Pi_1, \Pi_2)  \hfill &   - j (\Pi_1, \Pi_2, \Pi_3) \hfill  \\
        - \omega^{-2} j (\Pi_1, \Pi_2, \Pi_3)^* \hfill &  0                    \hfill  \\
   \end{array} }} \right]   
\label{hat(J) (hat(Pi_1), hat(Pi_2), hat(Pi_3) =}
\end{equation}

\subsection{Hodge operator}
Let us consider the oriented Euclidean space $\mathcal{T} = \mathbb{R}^n$ of  positive index of inertia $p$. Owing (\ref{U^* = U^T G}), the skew-adjoint map $M$ can be identified to the 2-form $A_M$ through
$$ U^* M \, V = U^T A_M V = A_M (U,V) \quad \mbox{with}  \quad 
   A_M = G \, M
$$
Then $M = \mathcal{J} (V_1, \ldots, V_{n - 2})$ can be identified to the 2-form $A_M = - vol (V_1, \ldots, V_{n - 2})$.  The vector space $\bigwedge^q \mathcal{T}$ of $q$-forms is an Euclidean space for the scalar product 
$$ G_q (A, B) = \frac{1}{q!} \, \sum_{\begin{array}{cc}
       i_1 \ldots i_q  \\
       j_1 \ldots j_q  \\
   \end{array}}
   G^{i_1 j_1} \ldots G^{i_q j_q} \,
   A_{i_1 \ldots i_q} B_{j_1 \ldots j_q }
$$
The {\bf adjoint}\footnote{To do not confuse with the adjoint of a linear map.} of a $q$-form $A$ is the $(n - q)$-form $* A$ such that
$$ (* A)  (V_1, \ldots, V_{n - q}) = (-1)^{q \, (n - q)}
   G_{q} (A, vol (V_1, \ldots, V_{n - q}))
$$
The linear map $* : \bigwedge^q \mathcal{T} \rightarrow \bigwedge^{n - q} \mathcal{T} : A \mapsto * A$ is called \textbf{Hodge operator}.
The following result can be proved:

\begin{theorem}
\label{theorem - Properties of the Hodge operator}
Properties of the Hodge operator:
\begin{itemize}
    \item [$\diamondsuit$] $M = \mathcal{J} (V_1, \ldots, V_{n - 2})$ is identified to the 2-form  
    $$ A_M = * (V^*_1 \wedge \ldots \wedge V^*_{n - 2}) = - vol (V_1, \ldots, V_{n - 2})
    $$ 
    \item  [$\heartsuit$] If $A$ is a $q$-form, $ * (* A) = (-1)^{q (n - 1) + n - p} \, A$ 
    \item  [$\spadesuit$] $* vol (V_1, \ldots, V_{n - 2}) = (-1)^{n - p} \, V^*_1 \wedge \ldots \wedge V^*_{n - 2} $
    \item[$\natural$] $* (V^* \wedge U^*) = - vol (U, V)$
\end{itemize}
\end{theorem}

In the special case of interest : $n = 4, q = 2$, if $A_M$ is a 2-form, $* A_M$ is also a 2-form identified to an skew-adjoint map denoted $*M$. If $V^*_1 \wedge V^*_2 = A_M'$, then $M' = G^{-1} (V^*_1 \otimes V^*_2 - V^*_2 \otimes V^*_1) = V_1 V^*_2 - V_2 V^*_1$. On the other hand, $\diamondsuit$ reads $A_{\mathcal{J} (V_1,V_2)} = * A_M' = A_{*M'}$ then:
\begin{equation}
   * ( V_1 V^*_2 - V_2 V^*_1)  = \mathcal{J} (V_1,V_2)
\label{* ( V_1 V^*_2 - V_2 V^*_1)  = J (V_1,V_2)}   
\end{equation}
that implies, owing to $\heartsuit$
\begin{equation}
   * \mathcal{J} (V_1,V_2) = (-1)^{n - p} (V_1 V^*_2 - V_2 V^*_1)   
\label{* J (V_1,V_2) = (-1)^(n - p) (V_1 V^*_2 - V_2 V^*_1)}
\end{equation}

\section{Coadjoint orbit method to classify the elementary particles}

This Section is a simple reminder of concepts of symplectic mechanics. For more details on this topics, the reader can consult for instance \cite{SSD, Abraham, Guillemin Sternberg 1984, Libermann Marle, SSDEng}.
In the sequel, all the considered Lie groups are matrix groups.
Any Lie group $G$ left linearly acts on its Lie algebra $\mathfrak{g}$ by the adjoint representation 
\begin{equation}
Ad \left( a \right)  : \mathfrak{g} \rightarrow \mathfrak{g} : Z' \mapsto Z = a Z' a^{-1} 
\label{Ad}
\end{equation}
$G$ left linearly acts on the dual $\mathfrak{g}^{*} $ of $\mathfrak{g}$ by the coadjoint representation $Ad^{*}$ such that
\begin{equation}
\forall Z \in \mathfrak{g}, \forall \mu \in \mathfrak{g}^{*}, \qquad
  \left(  Ad^{*} (a) \mu \right) \left( Z \right)  =  \mu \left(  Ad \left( a^{-1} \right)  Z \right)  
\label{Ad*}
\end{equation}

Let $\left( \mathcal{M},\varpi \right)$ be a symplectic manifold, an action of a Lie group $G$ on $\mathcal{M}$ 
$$ G \times \mathcal{M} \rightarrow \mathcal{M} : \left( a,x \right) \mapsto x' = a \cdot x $$ 
which is symplectic
$$   L^*_a \varpi = \varpi
$$
and a  momentum mapping $\psi : \mathcal{M} \rightarrow \mathfrak{g}^{*}$
$$\forall Z \in \mathfrak{g}, \, \forall dx \in T_{x} \mathcal{M}, \quad 
                   \varpi \left( Z \cdot x , dx \right)  = - d \left( \psi \left( x \right) Z \right)  $$
If the manifold $\mathcal{M}$ is connected, two momentum mappings differ by a constant. All the symplectic manifolds considered in the sequel are assumed to be connected. We consider now $\mathfrak{g}^{*}$ as equipped with the associated structure of affine space. The momentum mappings belong to the affine space of the mappings from $\mathcal{M}$ into  $\mathfrak{g}^{*}$, called the momentum mapping space. Any element $A$ of the affine group $\mathbb{GA} \left(  \mathfrak{g}^{*} \right) $ of $\mathfrak{g}^{*}$ is of the form $A \left( \mu \right) = P \mu + \mu_{0} $ where $P \in \mathbb{ GL} \left(  \mathfrak{g}^{*} \right)  $ and $ \mu_{0} \in \mathfrak{g}^{*}$. The action of $G$ on $\mathcal{M}$ induces a smooth right action on the momentum mapping space
$$ \forall x \in \mathcal{M}, \forall a \in G, \quad 
               \left(  \psi \cdot a \right)  \left(  x \right)  = \psi \left(  a \cdot x \right) $$
\begin{theorem}[Souriau]

Let $\left( \mathcal{M}, \varpi \right)$ be a connected symplectic manifold and a symplectic action $ \left( a,x \right) \mapsto x' = a \cdot x $ of a Lie group $G$. Then, the induced action $ \left( a, \psi \right) \mapsto \psi' = \psi \cdot a $ has the form 
\begin{equation}
\psi  \cdot a = Ad^{*} \left( a \right)  \psi + cocs \left(  a \right) 
\label{induce action}
\end{equation} where $cocs \left(  a \right) $ does not depends on $x$. 
The mapping $ cocs : G \rightarrow \mathfrak{g}^{*}$ is called a  symplectic cocycle.
\label{thm JMS}
\end{theorem}

\vspace{0.5cm}
The demonstration can be found for instance in Theorem (11.17) of \cite{SSD, SSDEng} and Section 4.2 of \cite{Abraham}. In the sequel, we consider only symmetry groups for which the consideration of symplectic cocycles is not relevant, then the group acts linearly.

The symplectic structure of the coadjoint  orbits is revealed by the following result.
\vspace{0.3cm}

\begin{theorem}[Kirillov-Kostant-Souriau]

Let $G$ be a Lie group and an orbit of the coadjoint representation $orb\,(\mu )\subset \mathfrak{g}^*$. Then:
\begin{itemize}
\item [$\diamondsuit$] The inclusion map $orb\,(\mu )\to \mathfrak{g}^* $ is a regular embedding. A vector $d \mu \in T_\mu \mathfrak{g}^* $ is tangent to the orbit if there exists $Z_d \in \mathfrak{g}$ such that: 
$$ d\,\mu =   \mu \,\circ \,ad\,(Z_d ) 
          = - ad^ *(Z_d ) \mu          
$$
\item [$\heartsuit$] The orbit $orb\,(\mu )$ is a symplectic manifold of which the symplectic form is defined by:
$$ \varpi_{KKS} (d \mu ,\delta\mu ) = \mu \,\left[ {Z_d ,Z_\delta }\right] 
$$
The dimension of the orbit is even.
\item [$\spadesuit$] $G$ is a symplectic group and any $\mu \in \mathfrak{g}^*$ is its own momentum.
\end{itemize}
\label{KKS thm} 
\end{theorem}

The reader can find a demonstration in Theorem (11.34) of \cite{SSD, SSDEng}. 
\vspace{0.5cm}


In physical terms, every observer is working in a reference frame. 
The symmetries $a \in G$ are changes of reference frames  or, equivalently,  of observers.
According to the mechanistic description of elementary particles proposed by Souriau, an isolated dynamical 
system is said to be elementary when the symmetry group acts transitively on the space of its motions, \textit{i.e.} it is the set of all possible motions of the system. The momentum map of its action is then a symplectic diffeomorphism of this space onto a coadjoint orbit of the group. For him, the so defined elementary systems are mathematical  models for elementary particles of physicists.

\section{Elementary particles in Poincar\'e relativity}
\label{Section Elementary particles in Poincare relativity}

In special relativity, the Universe $\mathcal{U}$ is represented by a 4D hyperbolic Euclidean space\footnote{In general relativity, the Universe is a manifold and we have to consider the tangent space regarding  to the elementary particles.}. An event $\bm{X} \in \mathcal{U}$ occuring in a reference frame at time $t$ and position $x \in \mathbb{R}^3$ is represented  by its coordinates
$$ X = \left[ {{\begin{array}{cc}
       t \hfill  \\
       x  \hfill  \\
   \end{array} }} \right]
$$
$\mathcal{U}$ is equipped with the Minkowski 1 + 3 metric represented in an orthonormal basis by Gram's matrix 
\begin{equation}
 G = 
\left[ {{\begin{array}{cc}
       1 \hfill &  0 \hfill \\
       0 \hfill &  - 1_{\mathbb{R}^3} \hfill \\
   \end{array} }} \right]
\label{G =}
\end{equation}
then in this basis the speed of the light $c$ is equal to 1.
The Lorentz transformations are the linear transformations $P$ of $\mathbb{R}^4$ which conserve the 1 + 3 metric, then such 
$$ P^* P = 1_{\mathbb{R}^4}
$$
In terms of the boost (or velocity of transport) $v \in \mathbb{R}^3$ and a rotation $R \in \mathbb{SO}(3)$, using the factor $\gamma = 1 / \sqrt{1 - \parallel v \parallel^2}$, the structure of a Lorentz transformation is 
$$ P = 
   \left[ {{\begin{array}{cc}
         \gamma \hfill &  \gamma \, v^T R \hfill \\
        \gamma \, v \hfill 
        & \left[1_{\mathbb{R}^4} + \frac{\gamma^2}{\gamma + 1} \, v \, v^T\right] \, R \hfill \\
   \end{array} }} \right]
\label{Lorentz transformation}
$$
In the sequel, we consider only the special Lorentz transformations \textit{i.e.} that belong to the connected component of the identity $\mathbb{SO}^+(1,3)$, a Lie group of dimension 6. The Poincar\'e group $\mathbb{G}$ is the set of affine transformations of $\mathbb{R}^4$ 
\begin{equation}
     X' \mapsto X = C + P \, X'
\label{X' mapsto X = C + P X'}
\end{equation}
of which the linear part $P$ is a Lorentz transformation and that we denote $a = (C, P)$. The translation is decomposed into a clock change $\tau$ and a space translation $k$
$$ C = \left[ {{\begin{array}{cc}
       \tau \hfill  \\
       k  \hfill  \\
   \end{array} }} \right]
$$
$\mathbb{G}$  is a Lie group of dimension 10. The elements of its Lie algebra $\mathfrak{g}$ are characterized by 
$$ Z \in \mathfrak{g} \quad \Leftrightarrow \quad 
Z = \delta a = (\delta C, \delta P) \; \mbox{such that} \; \delta P \; 
\mbox{is skew-adjoint}
$$
then
\begin{equation}
     \delta C = \left[ {{\begin{array}{cc}
       \delta \tau \hfill  \\
       \delta k  \hfill  \\
   \end{array} }} \right], \qquad
   \delta P =
   \left[ {{\begin{array}{cc}
       0 \hfill &  \delta v^T \hfill \\
       \delta v \hfill &  j (\delta \theta) \hfill \\
   \end{array} }} \right]
\label{delta C = & delta P = for Poincare}
\end{equation}
where $\delta R = j (\delta \theta) $ is an infinitesimal rotation around the identity of $\mathbb{R}^3$.
The corresponding momenta $\mu$ are linear forms on $\mathfrak{g}$
\begin{equation}
    \mu (Z) = - \Pi^* \delta C  - \frac{1}{2} Tr (M \, \delta P)
\label{mu (Z) = - Pi delta C - (1/2) Tr (M delta P)}
\end{equation}
characterized by
$$ \mu \in \mathfrak{g}^* \quad \Leftrightarrow \quad 
\mu = (\Pi, M) \; \mbox{such that} \; M \; 
\mbox{is skew-adjoint}
$$
For physical purpose, we introduce the 1 + 3 block decomposition
\begin{equation}
     \Pi = \left[ {{\begin{array}{cc}
       m \hfill  \\
       p  \hfill  \\
   \end{array} }} \right], \qquad
   M =
   \left[ {{\begin{array}{cc}
       0 \hfill &  r^T \hfill \\
       r \hfill &  j (l) \hfill \\
   \end{array} }} \right]
\label{decomposition of Pi & M for Poincare}
\end{equation}
where $m \in \mathbb{R}$ is the mass (equal to the energy because $c = 1)$, $p, r, l \in \mathbb{R}^3$ are respectively the linear momentum, the passage and the angular momentum. Then (\ref{mu (Z) = - Pi delta C - (1/2) Tr (M delta P)}) itemizes into
$$ \mu (Z) = l \cdot \delta \theta - r \cdot \delta v + p \cdot \delta k - m \, \delta \tau
$$
The coadjoint representation reads
\begin{equation}
     \mu = Ad (a) \, \mu' \quad \Leftrightarrow \quad 
 \Pi = P \, \Pi', \quad 
 M = P \, M' P^* + C \, (P \, \Pi')^* - (P \, \Pi') \, C^*
\label{mu = Ad (a) mu'}
\end{equation}

In Chapter 3, \S 14 of \cite{SSD, SSDEng}, Souriau uses the type (timelike or lightlike) of the 4-momentum vector (or energy-momentum vector) $\Pi$ for a classification of elementary systems. By these means, he obtains a large part of the physicists' classification of elementary particles. Below, briefly summarized, his results are presented. 

Let us suppose that $\Pi$ {\bf is timelike} and define the {\bf spin momentum}
\begin{equation}
    M_0 = M + \Pi \, X^* - X \, \Pi^*
\label{M_0 = M + Pi X^* - X Pi^* spin-momentum def}
\end{equation}
from which we deduce of (\ref{mu = Ad (a) mu'}) its transformation law
\begin{equation}
    M_0 = P \, M'_0 P^*
\label{transformation law of the spin M_0}
\end{equation}
and two properties 
\begin{itemize}
    \item [(a)] $M_0$ is skew-adjoint
    \item [(b)] The set of $X \in \mathbb{R}^4$ such that $M_0 \, \Pi = 0$ is a straight line $\mathcal{D}$ parallel to $\Pi$
\end{itemize}
Owing to (\ref{transformation law of the spin M_0}), $M_0$ does not depends on $X$ when $X$ runs over $\mathcal{D}$. Taking into account that a non null vector orthogonal to a timelike vector is spacelike (see \S 28 of \cite{CL}), we introduce $I, J \in \mathbb{R}^4$ such that 
\begin{equation}
    \Pi = m_0 \, I, \qquad I^* I = 1,\qquad J^* J = -1, \qquad J^* I = 0
\label{ Pi = m I & I^* I = 1 & J^* J = -1 & J^* I = 0}
\end{equation}
On this ground, we put
\begin{equation}
    M_0 = s \, \mathcal{J} (I, J)
\label{M_0 = s J (I, J)}
\end{equation}
verifying the properties (a) and (b) because of Theorem \ref{thm vector product} $\heartsuit$ and $\spadesuit$. The momentum $\mu = (\Pi, M)$ is characterized by $\mathcal{D}$  and $J$ or, equivalently, by $X, I, J$. 
Owing to (\ref{* ( V_1 V^*_2 - V_2 V^*_1)  = J (V_1,V_2)}), we have
$$ * M = * M_0 
     + * (X \, \Pi^* - \Pi \, X^*) = * M_0  + \mathcal{J} (X, \Pi)
$$
Let us introduce the \textbf{polarization}
\begin{equation}
     W = (* M) \, \Pi
\label{polarization W = (* M) Pi}
\end{equation}
Then, taking into account Theorem \ref{thm vector product} $\heartsuit$, (\ref{M_0 = s J (I, J)}) and  (\ref{* J (V_1,V_2) = (-1)^(n - p) (V_1 V^*_2 - V_2 V^*_1)})
$$ W = (* M_0) \, \Pi + \mathcal{J} (X, \Pi) \, \Pi 
   = s \, m_0 \, (* \mathcal{J} (X, \Pi)) \, I 
   =  s \, m_0 \, (J \, I^* - I \, J^*) \, I
$$
or, taking into account (\ref{ Pi = m I & I^* I = 1 & J^* J = -1 & J^* I = 0})
$$      W = s \, m_0 \, J
$$
The straight line of direction $J$ is called the {\bf polarization line}.

A {\bf particle with spin} is characterized by two non vanishing 4-vectors, its energy-momentum $\Pi$ (timelike) and its polarization $W$ (spacelike). The two numbers (Casimirs) 
$$ C_2 = \Pi^* \Pi > 0, \qquad C_4 = W^* W < 0  
$$
are invariant of the orbit, then do not depend on the reference frame or the observer. The calculus of the dimension of the isotropy group of the momentum $\mu$ shows that they are only 2 independent invariants, then every invariant is a combinations of the Casimirs. The numbers 
$$ m_0 = \sqrt{\Pi^* \Pi}, \qquad 
   s = \frac{\sqrt{- W^* W}}{\sqrt{\Pi^* \Pi}}
$$
are interpreted as the {\bf rest mass} and the \textbf{spin} of the particle. In particular, let 
$$  \Pi' = 
\left[ {{\begin{array}{cc}
       m_0 \hfill  \\
       0 \hfill  \\
   \end{array} }} \right]
$$
be the 4-momentum in the comoving frame (or proper reference frame). Applying to it a boost $v$, {\it i.e.} a Lorentz transformation (\ref{Lorentz transformation}) where the rotation is the identity, we obtain by (\ref{mu = Ad (a) mu'}) the mass and the linear momentum in the reference frame of any observer
$$ m = m_0 \gamma, \qquad p = m_0 \gamma \, v
$$
This relation can be expressed also as
\begin{equation}
 \Pi = m_0 \, U \quad \mbox{such that} \quad
   U^* U = 1
\label{Pi = m_0 U & U^* U = 1}    
\end{equation}
It is worth to remark that the straight line $\mathcal{D}$ is the trajectory of the particle (Chapter 3, \S 14 of \cite{SSD, SSDEng}). If it is parameterized by the arc length $s$ (for the  metric $G$), $U$ is the 4-velocity
\begin{equation}
     U = \frac{dX}{ds}  = 
    \left[ \begin{array}{c}
       \gamma \hfill  \\
       \gamma \, v \hfill  \\
   \end{array}  \right]
\label{U = dX / ds}    
\end{equation}

They are two other kind of orbits characterizing a {\bf particle without spin} and a {\bf massless particle}. For more details the reader is referred to (14.24) and (14.29) of \cite{SSD, SSDEng}.

We know that elementary particles are characterized by their mass, spin and electric charge. This last property is missing from this formalism and is usually reintroduced in the particle description as an external element to the geometric approach (see Chapter 3, \S 15 of \cite{SSD, SSDEng}). Our aim now is to develop a paradigm compatible with the experimental observations and in which the charge naturally appears as a momentum by constructing a bridge between the coadjoint orbit method and the Kaluza-Klein theory.

\section{Elementary particles in Kaluza-Klein relativity}
\label{Section - Elementary particles in Kaluza-Klein relativity}

In this theory, the Universe $\hat{\mathcal{U}}$ is represented by a 5D hyperbolic Euclidean space where the fifth dimension is curled up and microscopic. An event $\hat{\bm{X}} \in \hat{\mathcal{U}}$ occuring in a reference frame at time $t$, position $x$ and fifth coordinate $y$ is represented by the 5-column
$$ \hat{X} = \left[ {{\begin{array}{cc}
       X \hfill  \\
       y  \hfill  \\
   \end{array} }} \right]
   = \left[ {{\begin{array}{cc}
       t \hfill  \\
       x  \hfill  \\
       y  \hfill  \\
   \end{array} }} \right]
$$
$\hat{\mathcal{U}}$ is equipped with a 1 + 4 metric represented in an orthonormal basis by Gram's matrix 
\begin{equation}
 \hat{G} = 
\left[ {{\begin{array}{cc}
       G \hfill &  0 \hfill \\
       0 \hfill &  - 1 \hfill \\
   \end{array} }} \right]
\label{hat(G) =}
\end{equation}
where $G$ is given by (\ref{G =}).
The set $\hat{\mathbb{G}}_1$ of affine transformations $\hat{a} = (\hat{C}, \hat{P})$ of $\mathbb{R}^5$ of which the linear part conserves the metric 
$$ \hat{P}^* \hat{P} = 1_{\mathbb{R}^4}
$$
is a Lie group of dimension 15. The structure of the transformations of $\hat{\mathbb{G}}_1$ is given by
$$ \hat{C} = \left[ {{\begin{array}{cc}
       C \hfill  \\
       \xi  \hfill  \\
   \end{array} }} \right], \qquad
\hat{P} = 
\left[ {{\begin{array}{cc}
       P \hfill &  \beta \, P^{*-1} b  \hfill \\
       b^* \hfill &  \beta \hfill \\
   \end{array} }} \right]
$$
where 
\begin{eqnarray}
    b\in \mathbb{R}^4, \quad\qquad \beta = \sqrt{1 + b^* b} 
    \qquad\qquad \qquad\qquad\qquad\qquad \qquad\qquad
    \quad\quad \nonumber\\
P = P_L B, \qquad P_L \;\; \mbox{is a Lorentz transformation},
\qquad B = 1_{\mathbb{R}^4} + \frac{1}{\beta + 1} \, b \, b^* 
\end{eqnarray}
Its restriction to $\mathbb{R}^4$ is Poincaré's group.  The elements of its Lie algebra $\hat{\mathfrak{g}}_1$ are characterized by 
$$ \hat{Z} \in \hat{\mathfrak{g}}_1 \quad \Leftrightarrow \quad 
\hat{Z} = \delta \hat{a} = (\delta \hat{C}, \delta \hat{P}) \; \mbox{such that} \; \delta \hat{P} \; 
\mbox{is skew-adjoint}
$$
then
$$ \delta \hat{C} = \left[ {{\begin{array}{cc}
       \delta C \hfill  \\
       \delta \xi  \hfill  \\
   \end{array} }} \right], \qquad
   \delta \hat{P} =
   \left[ {{\begin{array}{cc}
       \delta P \hfill &  \delta b \hfill \\
       \delta b^* \hfill &  0 \hfill \\
   \end{array} }} \right]
$$
where $\delta P $ is skew-adjoint with respect to the metric (\ref{G =}).

Following Souriau, the momenta of the elementary particles can be obtained by considering the group action on the coadjoint orbits.
The momenta $\hat{\mu}$ of $\hat{\mathbb{G}}_1$ are of the form
\begin{equation}
    \hat{\mu} (\hat{Z}) = - \hat{\Pi}^*  \delta \hat{C}  - \frac{1}{2} Tr (\hat{M} \, \delta \hat{P})
\label{hat(mu) (hat(Z)) = - hat(Pi) delta hat(C) - (1/2) Tr (hat(M) delta hat(P)}
\end{equation}
characterized by
$$ \hat{\mu} \in \hat{\mathfrak{g}}^*_1 \quad \Leftrightarrow \quad 
\hat{\mu} = (\hat{\Pi}, \hat{M}) \; \mbox{such that} \; \hat{M} \; 
\mbox{is skew-adjoint}
$$
For convenience, we decompose the momenta as follows
\begin{equation}
 \hat{\Pi} = \left[ {{\begin{array}{cc}
       \Pi \hfill  \\
       q  \hfill  \\
   \end{array} }} \right], \qquad
   \hat{M} =
   \left[ {{\begin{array}{cc}
       M \hfill &  Q \hfill \\
       Q^* \hfill &  0 \hfill \\
   \end{array} }} \right]
\label{hat(Pi) = (Pi, q) & hat(M) = ((M, Q),(Q^* 0))}
\end{equation}
where the scalar $q$ and $Q \in \mathbb{R}^4$ are extra momenta with respect to those of Poincar\'e's group. The natural conjecture is to identify $q$ to the {\bf electric charge} with suitable physical units. This idea is mentioned in \cite{PETIT 2014} but the extension of Poincar\'e's group proposed by the authors is only a subgroup of $\hat{\mathbb{G}}_1$ of dimension 11. In the present work, the aim is to perform an in-depth investigation by examining the consequences of this conjecture to verify if they fit the observations. 

Then (\ref{hat(mu) (hat(Z)) = - hat(Pi) delta hat(C) - (1/2) 
Tr (hat(M) delta hat(P)}) itemizes into
\begin{equation}
     \hat{\mu} (\hat{Z}) = - \Pi^* \delta C  - \frac{1}{2} Tr (M \, \delta P) - q \, \delta \xi - Q^* \delta b
\label{hat(mu) (hat(Z)) = - Pi delta C - (1/2) Tr (M delta P) - q delta xi - Q^* delta b}
\end{equation}
The coadjoint representation reads
\begin{equation}
     \hat{\mu} = Ad (\hat{a})^* \, \hat{\mu}' \quad \Leftrightarrow \quad 
 \hat{\Pi} = \hat{P} \, \hat{\Pi}', \quad 
 \hat{M} = \hat{P} \, \hat{M}' \hat{P}^* + \hat{C} \, (\hat{P} \, \hat{\Pi}')^* - (\hat{P} \, \hat{\Pi}') \, \hat{C}^*
\label{hat(mu) = Ad^* (hat(a)) hat(mu)' for G_1}
\end{equation}
The action of $\hat{\mathbb{G}}_1$ on the {\bf energy-momentum-charge vector} $\hat{\Pi}$ itemizes into
$$ \Pi = P \, \Pi' + q' \, \beta \, P^{*-1} b, \qquad 
q = b^* \Pi + \beta \, q'
$$
It results that the electric charge is dependent on the reference frame then on the observer, totally at odds with the experimental observations. Likewise for Poincar\'e's group, we assume that $\hat{\Pi}$ is timelike. The only invariant is 
$$ \hat{\Pi}^* \hat{\Pi} = \Pi^* \Pi - q^2  > 0
$$
The value of the charge depends on the observer and may even be zero in certain reference frames. The charge is not characteristic only of the particle.
Author's opinion is that $\hat{\mathbb{G}}_1$ is not the symmetry group of the Universe today as we know it but nonetheless must not be {\it a priori} rejected. We have only to find the Physics that could admit it as symmetry group. However before to discuss this point, our goal now is to find a more appropriate symmetry group for the Physics today. It is what we shall be going to see in the next Section. Latter, we shall come back to the analysis of the structure of the momenta $\hat{\mu}$ of $\hat{\mathbb{G}}_1$ and the exhaustive determination of the invariants of the motion.

\section{Elementary particles in an Universe with an overwhelmingly small fifth dimension}
\label{Section - Elementary particles in an Universe with a microscopic fifth dimension}

\subsection{Zoom in}
\label{Subsection Zoom in}

To avert the previous paradox concerning the electric charge, it is worth to remark that the fifth dimension is curled up and the estimate of the cylinder radius (about $10^{-31}$ cm) is overwhelmingly small with respect to usual lengths in nuclear physics. The idea is to take advantage of this fact by a zoom in along the fifth coordinate and to consider the limit when the cylinder radius $\omega$ vanishes. Then we consider the transformation law of a vector $\hat{\bm{V}}$
\begin{equation}
     \hat{V}' = \hat{P}^{-1}_\omega \hat{V}
\label{zoom in}
\end{equation}
with the scaling
\begin{equation}
     \hat{P}_\omega =
   \left[{{\begin{array}{cc}
       1_{\mathbb{R}^4} \hfill &  0 \hfill \\
       0 \hfill &  \omega \hfill \\
   \end{array} }} \right]
\label{hat(P)_omega =}
\end{equation}
in such way that if the length was of the order of the cylinder radius in the fifth coordinate, it is after scaling of the order of the unity in the new coordinate (with a prime). Owing to the transformation law of 2-covariant tensors,
$$  \hat{G}' = \hat{P}^{T}_\omega \hat{G} \, \hat{P}_\omega
$$
the metric and its inverse are represented after the scaling by
\begin{equation}
     \hat{G}' = 
\left[ {{\begin{array}{cc}
       G \hfill &  0 \hfill \\
       0 \hfill &  - \omega^2 \hfill \\
   \end{array} }} \right], \qquad
\hat{G}'^{-1} = 
\left[ {{\begin{array}{cc}
       G^{-1} \hfill &  0 \hfill \\
       0 \hfill &  - \omega^{-2} \hfill \\
   \end{array} }} \right]
\label{hat(G)' & hat(G)'^(-1)}
\end{equation}
Next we omit the primes and consider the limits of these two matrices when $\omega \rightarrow 0$
\begin{equation}
  \hat{G}_0 = \left[ {{\begin{array}{cc}
       G \hfill &  0 \hfill \\
       0 \hfill &  0 \hfill \\
   \end{array} }} \right], \qquad 
 - \omega^{-2} \, \left[ {{\begin{array}{cc}
       0 \hfill &  0 \hfill \\
       0 \hfill &  1 \hfill \\
   \end{array} }} \right] = - \omega^{-2} \, 
   \left[ {{\begin{array}{cc}
       0 \hfill  \\
       1  \hfill  \\
   \end{array} }} \right] \otimes 
   \left[ {{\begin{array}{cc}
       0 \hfill  \\
       1  \hfill  \\
   \end{array} }} \right] 
\label{hat(G)_0 =}
\end{equation}
The first matrix represents a covariant semi-metric of signature $(+---0)$ while the second one represents a contravariant semi-metric of signature $(0\, 0 \, 0 \, 0-)$. After scaling, the Euclidean structure of $\hat{\mathcal{U}}$ is canceled  but it remains two debris, a symmetric 2-covariant tensor that we denote $\hat{\bm{G}_0}$ represented by $\hat{G}_0$ and a symmetric 2-contravariant tensor or, equivalently, a vector $\hat{\bm{\Omega}}_0$ represented by the column 
\begin{equation}
 \hat{\Omega}_0 =
    \left[ {{\begin{array}{cc}
       0 \hfill  \\
       1  \hfill  \\
   \end{array} }} \right]
\label{hat(Omega)_0 =}
\end{equation}
The set $\hat{\mathbb{G}}_0$ of affine transformations $\hat{a} = (\hat{C}, \hat{P})$ of $\mathbb{R}^5$ of which the linear part conserves the components of $\hat{\bm{G}_0}$ and $\hat{\bm{\Omega}}_0$ is such that
\begin{equation}
      \hat{C} = \left[ {{\begin{array}{cc}
        C \hfill  \\
        \xi  \hfill  \\
   \end{array} }} \right], \qquad
\hat{P} = 
\left[ {{\begin{array}{cc}
       P \hfill &  0  \hfill \\
       b^* \hfill &  1 \hfill \\
   \end{array} }} \right]
\label{C & P = for hat(G)_0}
\end{equation}
where $\xi$ is a scalar, $C, b\in \mathbb{R}^4$ and $P$ is a Lorentz transformation. As $\hat{\mathbb{G}}_1$, it is a Lie group of dimension 15. Then the affine transformation $\hat{X}' \mapsto \hat{X} = \hat{C} + \hat{P} \, \hat{X}'$ itemizes into
\begin{equation}
    X = C + P \, X', \qquad y = \xi + y' +  b^* X'
\label{X = C + P X' & y = xi + y' + b^* X'}
\end{equation}

\subsection{Gauge transformation}

To understand the meaning of $b$, we change of point of view and work in the framework of the general relativity in which $\hat{\bm{V}}$ is a tangent vector to a the 5D manifold $\hat{\mathcal{U}}$. As in \cite{AffineMechBook}, we hope to determine whether the G-structure in Kobayashi sense (\cite{Kobayashi 1972}) is integrable by solving thanks to Frobenius method the PDE system
$$ \frac{\partial \hat{X}}{\partial \hat{X}'} = \hat{P}, \qquad 
    \hat{P} \in G
$$
For $\hat{\mathbb{G}}_1$, the manifold $\hat{\mathcal{U}}$ is Riemannian and the the $\hat{\mathbb{G}}_1$-structure is not integrable if $\hat{\mathcal{U}}$ is curved. If it is flat, the transition maps are of the form $\hat{X}' \mapsto \hat{X} = \hat{C} + \hat{P} \, \hat{X}'$. For $\hat{\mathbb{G}}_0$, the compatibility conditions of the system provides the equation 
$$ b = \mbox{grad}_X \, h 
$$
where $X' \mapsto h(X')$ is a real function, that by integration leads to the gauge transformation
$$ y = y' + h (X')
$$
The $\hat{\mathbb{G}}_0$-structure is integrable only if the manifold is flat. Otherwise, it is convenient as in Chapter VII, (41.21) of \cite{GR} to consider the transition maps of standard charts 
$$ \hat{X} = \hat{H} \, (\hat{X}') =
\left[ {{\begin{array}{cc}
       H (X') \hfill  \\
       y' + h (X')  \hfill  \\
   \end{array} }} \right] 
$$
In terms of 4-potential $A$ of the electromagnetism, the vector $b$ turns out to be a gauge transformation
$$ A' \mapsto A = A' + b
$$

\subsection{Coadjoint orbit}
\label{SubSection - Coadjoint orbit}

The elements of the Lie algebra $\hat{\mathfrak{g}}_0$ of $\hat{\mathbb{G}}_0$ are characterized by 
\begin{equation}
 \hat{Z} \in \hat{\mathfrak{g}}_0 \quad \Leftrightarrow \quad 
\left\lbrace \begin{array}{l}
        \hat{Z} = \delta \hat{a} = (\delta \hat{C}, \delta \hat{P}) \; \mbox{such that} \; 
        \\ 
        \\
         \delta \hat{C} = \left[ {{\begin{array}{cc}
       \delta C \hfill  \\
       \delta \xi  \hfill  \\
   \end{array} }} \right], \quad
   \delta \hat{P} =
   \left[ {{\begin{array}{cc}
       \delta P \hfill &  0   \hfill \\
       \delta b^* \hfill &  0 \hfill \\
   \end{array} }} \right],
         \\
         \\
         \delta P \; \mbox{is skew-adjoint with} 
         \\
         \mbox{respect to the metric} \; (\ref{G =})
         \\
   \end{array} 
   \right.    
\label{hat(Z) in hat(g)_0 equivalent to}
\end{equation}
The momenta $\hat{\mu}$ of $\hat{\mathbb{G}}_0$ are of the form
$$ \hat{\mu} (\hat{Z}) = - \Pi^* \delta C  - \frac{1}{2} Tr (M \, \delta P) - q \, \delta \xi - Q^* \delta b
$$
The coadjoint representation $\hat{\mu} = Ad (\hat{a})^* \, \hat{\mu}'$ is defined by the following transformation laws
\begin{equation}
   \Pi = P \, (\Pi' - q' \, b)
\label{Pi = P (Pi' - q' b)}
\end{equation}
\begin{equation}
   M = P \, M' P^* + C \, (P \, (\Pi' - q \, b))^* - (P \, (\Pi' - q \, b)) \, C^*
    + (P \, b) (P \, Q')^* - (P \, Q') (P \, b)^*
\label{M = for hat(G)_0}
\end{equation}
\begin{equation}
   q = q'
\label{q = q'}
\end{equation}
\begin{equation}
  Q = P \, Q' + q' C
\label{Q = P Q' + q' C}
\end{equation}
In contrast to what happened with $\hat{\mathbb{G}}_1$, {\bf the electric charge is independent on the reference frame then on the observer}. This include in particular the invariance with respect to the gauge transformation. Besides, owing to (\ref{X = C + P X' & y = xi + y' + b^* X'}), the transformation law (\ref{Q = P Q' + q' C}) is satisfied if we claim that
\begin{equation}
     Q = q \, X
\label{Q = q X}
\end{equation}
that provides {\bf the physical interpretation of the momentum} $Q$ {\bf as the product of the charge  and the space-time position}. We suggest to call it the {\bf 4-position-charge momentum}.
Let us remark also that, taking into account the transformation law of $\Pi$, the one of $M$ is simplified as follow
\begin{equation}
   M = P \, M' P^* + C \, \Pi^* -  \Pi \, C^*
    + (P \, b) (P \, Q')^* - (P \, Q') (P \, b)^*
\label{M = for hat(G)_0 BIS}
\end{equation}

To calculate the number of independent invariants of the momentum of a group $G$, we determine the isotropy group of $\hat{\mu}$ from which we deduce the dimension of the coadjoint orbit
$$ \mbox{dim} \, (\mbox{orb} (\hat{\mu})) 
 = \mbox{dim} \, G - \mbox{dim} (\mbox{iso} (\hat{\mu}))
$$
The number of independent invariant of the orbit is
$$ n_I = \mbox{dim} \,  \mathfrak{g} - \mbox{dim} (\mbox{orb} (\hat{\mu})) 
       = \mbox{dim} (\mbox{iso} (\hat{\mu}))
$$
Taking into account (\ref{Pi = P (Pi' - q' b)}), (\ref{M = for hat(G)_0 BIS} and (\ref{Q = P Q' + q' C}), to determine the isotropy group of $\hat{\mu}$, we need to solve with respect to the unknowns $C, P, b, \xi$ the system of equations
\begin{equation}
   \Pi = P \, (\Pi - q \, b)
\label{Pi = P (Pi - q b) for iso}
\end{equation}
\begin{equation}
   M = P \, M P^* + C \, (P \, (\Pi - q \, b))^* - (P \, (\Pi - q \, b)) \, C^*
    + (P \, b) (P \, Q)^* - (P \, Q) (P \, b)^*
\label{M = for hat(G)_0 for iso}
\end{equation}
\begin{equation}
  Q = P \, Q' + q' C
\label{Q = P Q' + q' C for iso}
\end{equation}
the condition (\ref{q = q'}) being useless because the charge is an obvious invariant. 

For a {\bf charged particle} ($q \neq 0$), let us show that {\bf the number of independent invariants is 3}. Indeed, (\ref{Pi = P (Pi - q b) for iso}) and (\ref{Q = P Q' + q' C for iso}) give
\begin{equation}
   b = \frac{1}{q} \, (\Pi - P^* \Pi), \quad 
   C = \frac{1}{q} \, (Q - P \, Q)
\label{b = (1/q) (Pi - P^*) & C = (1/q) (Q - P Q)}
\end{equation}

Introducing these expressions of $b$ and $C$ into (\ref{M = for hat(G)_0 for iso}) leads to
\begin{equation}
   M_0 = P \, M_0  P^*
\label{M_0 = P M_0 P^* for G_0}
\end{equation}
where, taking into account (\ref{Q = q X})
\begin{equation}
   M_0 = M + \frac{1}{q} \, (\Pi \, Q^* - Q \, \Pi^*)
       = M + \Pi \, X^* - X \, \Pi^*
\label{M_0 = M + (1/q) (Pi Q^* - Q Pi^*}
\end{equation}
can be interpreted as the \textbf{spin momentum}, by comparison to (\ref{M_0 = M + Pi X^* - X Pi^* spin-momentum def}). 
As in Poincar\'e relativity, the straight line $\mathcal{D}$ of equation $M_0 \, \Pi = 0$ is the trajectory of the particle, parallel to $\Pi$ given by, owing to (\ref{Pi = m_0 U & U^* U = 1}) and (\ref{U = dX / ds})
\begin{equation}
     \Pi = m_0 \, U \quad 
    \mbox{with} \quad U = \frac{dX}{ds} \quad
    \mbox{such that} \quad
    U^* U = 1
\label{Pi = m_0 with U = dX/ds such that U^* U = 1}
\end{equation}

First we consider a \textbf{charged particle with spin} ($q \neq 0, \Pi$ timelike, $M_0 \neq 0$). Equation (\ref{M_0 = P M_0 P^* for G_0})  being non linear with respect to $P$ then difficult to solve, we use an infinitesimal method by working with its differential version in terms of Lie algebra. 
Differentiating  it with respect to $P$ at the identity and taking into account that $\delta P$ is skew-adjoint, it holds
$$ \delta P \, M_0 - M_0 \delta P = 0
$$
Putting
$$     \delta P =
   \left[ {{\begin{array}{cc}
       0 \hfill &  \delta v^T \hfill \\
       \delta v \hfill &  j (\delta \theta) \hfill \\
   \end{array} }} \right], \qquad
      M_0 =
   \left[ {{\begin{array}{cc}
       0 \hfill &  r^T_0 \hfill \\
       r_0 \hfill &  j (l_0) \hfill \\
   \end{array} }} \right]
$$
we obtain the system of equation
$$ l_0 \times \delta \, v + r_0 \times \delta \, \theta = 0, \qquad
   r_0 \times \delta \, v - l_0 \times \delta \, \theta = 0
$$
In the general case ($l_0 \times r_0 \neq 0$), the solution is
$$ \delta \, \theta  = \lambda \, r_0 + \mu \, l_0, \qquad 
   \delta \, v = \mu \, r_0 - \lambda \, l_0
$$
where $\lambda$ and $\mu$ are two independent scalars of arbitrary values. Then the dimension of the isotropy group of $M_0$ is 2 because  the couple $(\delta \, \theta, \delta \, v)$ then $\delta  \, P$ is defined by these 2 independent parameters.  It is easy to verify that the dimension 2 is also valid for the particular case. The value of $b$ and $C$ are fixed by (\ref{b = (1/q) (Pi - P^*) & C = (1/q) (Q - P Q)}). The value of $\xi$ that does not occurs in the equations is free. Then the dimension of the isotropy group of $\hat{\mu}$ and then the number of independent invariants is 3.

For a \textbf{charged particle with spin}, a set of 3 independent invariants of $\hat{\mu}$ is the \textbf{electric charge}, the \textbf{rest mass} and the \textbf{spin} of the particle
$$ q, \qquad
   m_0 = \sqrt{\Pi^* \Pi}, \qquad 
   s = \frac{\sqrt{- W^* W}}{\sqrt{\Pi^* \Pi}}
$$
where the polarization $W$ is defined by (\ref{polarization W = (* M) Pi}).


For a \textbf{charged particle without spin} ($q \neq 0, \Pi$ timelike, $M_0 = 0$), equation (\ref{M_0 = P M_0 P^* for G_0}) is satisfied for every $P$. The number independent invariants is 7. The invariants are $q$ and the 6 independent components of $M_0$ which are null.

For a {\bf particle without charge} ($q = 0, Q = 0$), the coadjoint representation of 
$\hat{\mathbb{G}}_0$ is reduced to (\ref{mu = Ad (a) mu'}) and we recover the classification of elementary particles in Poincar\'e's relativity.

\vspace{0.3cm}


To author's knowledge, the Physics of the symmetry group $\hat{\mathbb{G}}_0$ has not been considered in the literature. By the way, there exists a group introduced by  \cite{Levy-Leblond 1965} and called by him Carroll's group\footnote{Lewis Carroll, the author of {\it Alice's Adventures in Wonderland}, although mathematician, had nothing to do with the creation of this group.}, of which the elements have the same mathematical structure as the ones of $\hat{\mathbb{G}}_0$. However the physical meaning of Carroll's group is very different from the one of $\hat{\mathbb{G}}_0$: Carroll's group linearly acts on a space of dimension 4 (not 5) and the dimension analogous to the fifth of $\hat{\mathbb{G}}_0$ is timelike (not spacelike). Carroll's group was constructed as a degeneracy of Poincar\'e's group (similar but distinct from Galileo's group). Despite of these discrepancies concerning the physical aspects, Carroll's group may be considered as a forunner of the group $\hat{\mathbb{G}}_0$. For more details on Carroll's group, the reader is referred to \cite{Duval 1985, Duval 1991, Duval 2014, Bergshoeff 2014, Bergshoeff 2023, Levy-Leblond 2023}. 
\section{A cosmological scenario for the evolution of elementary particle structure}
\label{Section - A cosmological scenario for the evolution of elementary particle structure}

One of the important problems in Kaluza-Klein theories is how to explain the large separation of the scale of our 3D space and that of the extra dimension.
This issue was addressed in \cite{Okada 1986} with a Robertson-Walker metric in our four dimensions and a $d$-dimensional sphere $S^d$ in the extra dimensions. For our concern, $d = 1$. Under certain assumptions, this approach predicts stable cosmological models for the extra dimension and, using Kasner solutions for approximations when $t \rightarrow 0$, a possible evolution of the scale factors $a$ for our 3D space and $\omega$ for the fifth dimension given by
$$ a(t) \cong a_0 \, t^{1/2}, \qquad 
   \omega (t) \cong \omega_0 \, t^{-1/6}
$$
Although the assumptions are affected by many uncertainties, it seems reasonable to expect a cosmological scenario in which:
\begin{itemize}
    \item the elementary particles of the early 5D Universe are classified from the momenta of the group $\hat{\mathbb{G}}_1$,
    \item next the three former space dimensions inflate quickly while the last one shrinks,
    \item leading to the 4D era in which as today the particles are characterized by the momenta of the group $\hat{\mathbb{G}}_0$. 
\end{itemize}
{\bf By this mechanism, the elementary particles can acquire electric charge as a by-product of the $4 + 1$ symmetry breaking of the Universe.}

To describe it, we consider the metric $\hat{\bm{G}}_\omega$ at scale $\omega$. We omit the prime for the scaled coordinates. Owing to (\ref{hat(G)' & hat(G)'^(-1)}), the metric is represented by Gram's matrix 
$$ \hat{G}_\omega =
   \left[ {{\begin{array}{cc}
       G \hfill &  0 \hfill \\
       0 \hfill &  - \omega^2 \hfill \\
   \end{array} }} \right]
$$
The set $\hat{\mathbb{G}}_\omega$ of affine transformations $\hat{a} = (\hat{C}, \hat{P})$ of $\mathbb{R}^5$ of which the linear part conserves the metric $\hat{G}_\omega$ is a Lie group of dimension 15. The structure of the transformations of $\hat{\mathbb{G}}_\omega$ is given by
$$ \hat{C} = \left[ {{\begin{array}{cc}
       C \hfill  \\
       \xi  \hfill  \\
   \end{array} }} \right], \qquad
\hat{P} = 
\left[ {{\begin{array}{cc}
       P \hfill &  \omega^2 \beta \, P^{*-1} b  \hfill \\
       b^* \hfill &  \beta \hfill \\
   \end{array} }} \right]
$$
where  
\begin{eqnarray}
    b\in \mathbb{R}^4, \quad\qquad 
    \beta = \sqrt{1 + \omega^2 b^* b} 
    \quad\qquad \qquad\qquad\qquad\qquad \qquad\qquad
    \quad\quad \nonumber\\
P = P_L B, \qquad P_L \;\; \mbox{is a Lorentz transformation},
\qquad B = 1_{\mathbb{R}^4} + \frac{\omega^2}{\beta + 1} \, b \, b^* 
\end{eqnarray}
We would like to classify the elementary particles with charge at scale $\omega > 0$. In particular, we shall obtain the classification for the early Universe ($\omega = 1$) of which the study was only sketched out in Section \ref{Section - Elementary particles in Kaluza-Klein relativity}. The other limit case of the Universe today ($\omega = 0)$ that is singular was already studied in Section \ref{Section - Elementary particles in an Universe with a microscopic fifth dimension}.

The elements $\hat{a} = (\delta \hat{C}, \delta \hat{P})$ of the Lie algebra $\hat{\mathfrak{g}}_\omega$ of  $\hat{\mathbb{G}}_\omega$ are such that
\begin{equation}
     \delta \hat{C} = \left[ {{\begin{array}{cc}
       \delta C \hfill  \\
       \delta \xi  \hfill  \\
   \end{array} }} \right], \qquad
   \delta \hat{P} =
   \left[ {{\begin{array}{cc}
       \delta P \hfill &  \omega^2 \delta b \hfill \\
       \delta b^* \hfill &  0 \hfill \\
   \end{array} }} \right]
\label{delta hat(C) = & delta hat(P) = for G_omega}
\end{equation}
where $\delta P $ is skew-adjoint with respect to the metric (\ref{G =}). The momenta $\hat{\mu}$ of $\hat{\mathbb{G}}_\omega$ are of the form (\ref{hat(mu) (hat(Z)) = - hat(Pi) delta hat(C) - (1/2) Tr (hat(M) delta hat(P)}).
We decompose the momenta as follows
\begin{equation}
     \hat{\Pi} = \left[ {{\begin{array}{cc}
       \Pi \hfill  \\
       q  \hfill  \\
   \end{array} }} \right], \qquad
   \hat{M} =
   \left[ {{\begin{array}{cc}
       M \hfill               &  Q \hfill \\
       \omega^{-2} Q^* \hfill &  0 \hfill \\
   \end{array} }} \right]
\label{hat(Pi) = & hat(M) = for G_omega}
\end{equation}
transforming (\ref{hat(mu) (hat(Z)) = - hat(Pi) delta hat(C) - (1/2) 
Tr (hat(M) delta hat(P)}) into (\ref{hat(mu) (hat(Z)) = - Pi delta C - (1/2) Tr (M delta P) - q delta xi - Q^* delta b}).
The coadjoint representation is defined by (\ref{hat(mu) = Ad^* (hat(a)) hat(mu)' for G_1}) or, in details
\begin{equation}
    \Pi = P \, \Pi' + \omega^2 q' \, \beta \, P^{*-1} b
\label{Pi = for Ad^* of G_omega}
\end{equation}
\begin{eqnarray}
        M & = & P \, M' P^* 
     + \beta \,\left[ (P^{*-1} b) (P \, Q')^* 
     -                (P \, Q') (P^{*-1} b)^* \right] \nonumber\\
 & & + \, C\, (P \, \Pi' + \omega^2 q' \, \beta \, P^{*-1} b)^*
     - (P \, \Pi' + \omega^2 q' \, \beta \, P^{*-1} b) \, C^*
\label{M = = for Ad^* of G_omega}
\end{eqnarray}
\begin{equation}
    q = b^* \Pi + \beta \, q'
\label{q = for Ad^* of G_omega}
\end{equation}
\begin{eqnarray}
        Q & = & \beta \, P \, Q' 
       - \omega^2 \,\left[ P \, M' b
                   + \beta \, (b^* Q') \, P^{*-1} b \right] \nonumber\\
   & & + \, \omega^2  \,\left[ 
   \xi \, (P \, \Pi' + \omega^2 q' \, \beta \, P^{*-1} b)      - (b^* \Pi' + \beta \, q' ) \, C \right]    
\label{Q = = for Ad^* of G_omega}
\end{eqnarray}

By analogy with (\ref{M_0 = M + Pi X^* - X Pi^* spin-momentum def}), we define the spin momentum
$$     \hat{M}_0 = \hat{M} + \hat{\Pi} \, \hat{X}^* - \hat{X} \, \hat{\Pi}^*
$$
from which we deduce of (\ref{hat(mu) = Ad^* (hat(a)) hat(mu)' for G_1}) its transformation law
\begin{equation}
    \hat{M}_0 = \hat{P} \, \hat{M}'_0 \hat{P}^*
\label{transformation law of the spin M_0 for G_1}
\end{equation}
and two properties 
\begin{itemize}
    \item [(i)] $\hat{M}_0$ is skew-adjoint
    \item [(ii)] The set of $\hat{X} \in \mathbb{R}^5$ such that $\hat{M}_0 \, \hat{\Pi} = 0$ is a straight line $\hat{\mathcal{D}}$ parallel to $\hat{\Pi}$
\end{itemize}

\subsection{Polarization plane and map}

At this stage, it is worth to remark an important difference between Poincar\'e relativity, in which $A_{M_0}$ and its  adjoint $* A_{M_0}$ are both 2-forms, and Kaluza-Klein one for which $A_{\hat{M}_0}$ is a 3-form while $* A_{\hat{M}_0}$ is a 2-form, that has important consequences from the geometrical and physical viewpoints. 
Owing to (\ref{transformation law of the spin M_0 for G_1}), $\hat{M}_0$ does not depends on $\hat{X}$ when $\hat{X}$ varies over $\hat{\mathcal{D}}$. Let $\hat{I}, \hat{J}_1, \hat{J}_2 \in \mathbb{R}^5$ such that 
\begin{equation}
    \hat{\Pi} = m_0 \, \hat{I}, \qquad \hat{I}^* \hat{I} = 1,\qquad \hat{J}^*_1 \hat{J}_1 =  \hat{J}^*_2 \hat{J}_2 = -1, \qquad
    \hat{J}^*_1 \hat{I} = \hat{J}^*_2 \hat{I} = 
    \hat{J}^*_1 \hat{J}_2 = 0
\label{ Pi = m I & I^* I = 1 & J^*_1 J_1 = -1 & J^*_1 I = 0}
\end{equation}
Over $\hat{\mathcal{D}}$, we put
\begin{equation}
    \hat{M}_0 = s \, \mathcal{J} (\hat{I},  \hat{J}_1, \hat{J}_2)
\label{hat(M)_0 = s J (hat(I), hat(J)_1, hat(J)_2)}
\end{equation}
verifying the properties (i) and (ii) because of Theorem \ref{thm vector product} $\heartsuit$ and $\spadesuit$. 
The momentum $\hat{\mu} = (\hat{\Pi}, \hat{M})$ is characterized by $\mathcal{D}$  and $\hat{J}_1, \hat{J}_2$ or, equivalently, by $\hat{X}, \hat{I}, \hat{J}_1, \hat{J}_2$. In Poincar\'e relativity, there is a polarization straight line of direction $J$ while in Kaluza-Klein relativity, there is a \textbf{polarization plane} spanned by the couple $(\hat{J}_1, \hat{J}_2)$ that is one of its orthonormal basis.
Owing to Theorem \ref{theorem - Properties of the Hodge operator} $\natural$, we have
$$ * A_{\hat{M}} = * A_{\hat{M}_0} 
     + * (\hat{X}^*  \wedge \hat{\Pi}^*)
     = * A_{\hat{M}_0}   
     + vol (\hat{\Pi}, \hat{X})
$$
then
$$ * A_{\hat{M}} (\hat{\Pi}, \hat{J}_k) = * A_{\hat{M}_0} (\hat{\Pi}, \hat{J}_k)
     + vol (\hat{\Pi}, \hat{X}, \hat{\Pi}, \hat{J}_k)
     = s \,* A_{\hat{\Omega}} (\hat{\Pi}, \hat{J}_k)
     \qquad (k = 1,2)
$$
that leads to investigate the properties of the 2-form $* A_{\hat{\Omega}}$. Applying Theorem \ref{thm vector product} $\heartsuit$ and $\spadesuit$, its adjoint is the 3-form
$$ * A_{\hat{\Omega}} = - * vol (\hat{I}, \hat{J}_1, \hat{J}_2) = - \hat{I}^* \wedge \hat{J}^*_1 \wedge\hat{J}^*_2
$$
from which we deduce 
\begin{equation}
\hat{J}_1 = - (* A_{\hat{\Omega}} (\hat{I}, \hat{J}_2))^* , \qquad
\hat{J}_2 =  (* A_{\hat{\Omega}} (\hat{I}, \hat{J}_1))^* 
\label{hat(J)_1 = & hat(J)_2 =}
\end{equation}
In terms of the momentum $\hat{\mu} = (\hat{\Pi}, \hat{M})$, introducing the \textbf{polarization map} 
$$ \mbox{pol}_{\hat{\mu}} : \mathbb{R}^5 \rightarrow \mathbb{R}^5 :
\hat{V} \mapsto \hat{J} = \mbox{pol}_{\hat{\mu}} (\hat{V} )
= (* A_{\hat{M}} (\hat{\Pi}, \hat{V}))^*
$$
the relations (\ref{hat(J)_1 = & hat(J)_2 =}) are recast as
$$ s \, m_0 \, \hat{J}_1 = - \mbox{pol}_{\hat{\mu}} (\hat{J}_2) , \qquad
s \, m_0 \, \hat{J}_2 = \mbox{pol}_{\hat{\mu}}  (\hat{J}_1)
$$
Now, we have to address the issues of the existence and (non) uniqueness of the orthonormal basis $(\hat{J}_1, \hat{J}_2)$ to define the polarization map: 
\begin{itemize}
    \item \textbf{Existence.} Let us $\hat{K}_1, \hat{K}_2$ two vectors of $\mathbb{R}^5$ that form with $\hat{I}, \hat{J}_1, \hat{J}_2$ an orthonormal basis of $\mathbb{R}^5$ in which we decompose any vector
    $$ \hat{U} = \lambda  \hat{I} + \mu_1 \hat{J}_1 + \mu_2 \hat{J}_2 + \nu_1 \hat{K}_1 + \nu_2 \hat{K}_2
    $$
    Then
    $$ \mbox{pol}_{\hat{\mu}} (\hat{U} ) = s \, m_0 \, (\mu_1 \hat{J}_2 - \mu_2 \hat{J}_1)
    $$
    that shows the image of the polarization map $\mbox{pol}_{\hat{\mu}}$ is the polarization plane $\mathcal{P}$. This gives a method to determine a basis of $\mathcal{P}$. We span $\mathbb{R}^5$ and apply the polarization map to find two linearly independent vectors. They form a basis of $\mathcal{P}$. Then we find an orthonormal basis by using the Gram–Schmidt process.
    \item \textbf{Non uniqueness.} Let $(\hat{J}_1, \hat{J}_2)$ and $(\hat{J}'_1, \hat{J}'_2)$ be two  orthonormal bases such that
    $$ \mathcal{J} (\hat{I},  \hat{J}_1, \hat{J}_2) = \mathcal{J} (\hat{I},  \hat{J}'_1, \hat{J}'_2)
    $$
    Owing to Theorem \ref{thm vector product} $\heartsuit$,
    $$ \mathcal{J} (\hat{I},  \hat{J}_1, \hat{J}_2) \, \hat{J}'_k = \mathcal{J} (\hat{I},  \hat{J}'_1, \hat{J}'_2) \, \hat{J}'_k = 0 \qquad (k = 1,2)
    $$
    then $\hat{J}'_1$ and $\hat{J}'_2$ are linear combinations of $\hat{J}_1$ and $\hat{J}_2$. This ensures the uniqueness of the polarization plane. In contrast, the orthonormal basis $(\hat{J}_1, \hat{J}_2)$ is defined only  modulo an orthogonal transformation of $\mathcal{P}$.
\end{itemize}

\subsection{Invariants of the motion}

For a {\bf charged particle} ($q \neq 0$), let us verify on a simple case that {\bf the number of independent invariants is 2}. Taking into account (\ref{hat(mu) = Ad^* (hat(a)) hat(mu)' for G_1}), we  determine the isotropy group of the momentum by solving the system of equations
$$ \hat{\Pi} = P \, \hat{\Pi}, \qquad 
 \hat{M} = \hat{P} \, \hat{M} \hat{P}^* + \hat{C} \, (\hat{P} \, \hat{\Pi})^* - (\hat{P} \, \hat{\Pi}) \, \hat{C}^*
$$
with respect to $\hat{a}= (\hat{C}, \hat{P})$. Like in Section \ref{SubSection - Coadjoint orbit}, we differentiate these equations with respect to $\hat{a}$ at the identity and we take into
account that $\hat{M}$ is skew-adjoint, that leads to solve the linear system
\begin{equation}
     \delta \hat{P} \, \hat{\Pi} = 0, \qquad
   \delta \hat{P} \, \hat{M} - \hat{M} \, \delta \hat{P} 
   + \delta \hat{C} \, \hat{\Pi}^* 
   - \hat{\Pi} \, \delta \hat{C}^* = 0
\label{delta hat(P) hat(Pi) = 0 & delta hat(P) hat(M) - hat(M) delta hat(P) + delta hat(C) Pi^* - Pi delta hat(C)^* = 0}
\end{equation}
with respect to $\delta \hat{a}= (\delta \hat{C}, \delta \hat{P})$. Let us pick up 
$$ \hat{I} = \left[ {{\begin{array}{cc}
       I \hfill  \\
       \kappa  \hfill  \\
   \end{array} }} \right]
   = \left[ {{\begin{array}{cc}
       \mu \hfill  \\
       0  \hfill  \\
       \kappa  \hfill  \\
   \end{array} }} \right], \qquad
   \hat{J}_k = \left[ {{\begin{array}{cc}
       J_k \hfill  \\
       0  \hfill  \\
   \end{array} }} \right]
   = \left[ {{\begin{array}{cc}
       0 \hfill  \\
       p_k  \hfill  \\
       0  \hfill  \\
   \end{array} }} \right], \qquad (k = 1, 2)
$$
that satisfy conditions (\ref{ Pi = m I & I^* I = 1 & J^*_1 J_1 = -1 & J^*_1 I = 0}). 
Owing to (\ref{delta hat(C) = & delta hat(P) = for G_omega}) and (\ref{hat(Pi) = & hat(M) = for G_omega}), the first equation of (\ref{delta hat(P) hat(Pi) = 0 & delta hat(P) hat(M) - hat(M) delta hat(P) + delta hat(C) Pi^* - Pi delta hat(C)^* = 0}) itemizes into
\begin{equation}
    \delta P \, \Pi + \omega^2 q \, \delta b = 0, \qquad 
    \delta b^* \Pi = 0
\label{delta P Pi + omega^2 q delta b = 0 & b^* Pi = 0}
\end{equation}
of which the solution is, owing to (\ref{delta C = & delta P = for Poincare}) and (\ref{decomposition of Pi & M for Poincare})
\begin{equation}
     \delta b = \left[ {{\begin{array}{cc}
       \delta b_0 \hfill  \\
       \delta \bar{b}  \hfill  \\
   \end{array} }} \right] =
    \left[ {{\begin{array}{cc}
       0 \hfill  \\
       - \frac{\mu}{\omega^2 \kappa} \, \delta v \hfill  \\
   \end{array} }} \right]
\label{delta b =}
\end{equation}
Besides, owing to (\ref{J (Pi_1, Pi_2, Pi_3) =}) and (\ref{J (Pi_1, Pi_2) =}), we have
$$ \mathcal{J} (I, J_1, J_2) = 
   \left[ {{\begin{array}{cc}
       0 \hfill  \\
       \mu \, \psi  \hfill  \\
   \end{array} }} \right], \qquad
   \mathcal{J} (J_1, J_2) = 
     \left[ {{\begin{array}{cc}
       0 \hfill & \psi^T\hfill  \\
       \psi  \hfill & 0 \hfill  \\
   \end{array} }} \right]
$$
with $\psi = p_1 \times p_2$, from which we deduce, taking into account (\ref{hat(M)_0 = s J (hat(I), hat(J)_1, hat(J)_2)}) and (\ref{hat(J) (hat(Pi_1), hat(Pi_2), hat(Pi_3) =})
$$ \hat{M}_0 = s \hat{\mathcal{J}} (\hat{I}, \hat{J}_1, \hat{J}_2) = 
 s \, \left[ {{\begin{array}{cc}
       \kappa \, \mathcal{J} (J_1, J_2) \hfill & - \mathcal{J} (I,J_1, J_2)\hfill  \\
       - \omega^2 \mathcal{J} (I,J_1, J_2)^*  \hfill & 0 \hfill  \\
   \end{array} }} \right]
$$
$$ \hat{M}_0 =
 s \, \left[ {{\begin{array}{ccc}
       0 \hfill & \kappa \, \psi^T \hfill & 0 \hfill  \\
       \kappa \, \psi  \hfill & 0 \hfill & - \mu \, \psi \hfill  \\
       0 \hfill & \mu \omega^{-2} \psi^T & 0 \hfill \\
   \end{array} }} \right]
$$
the second equation of (\ref{delta hat(P) hat(Pi) = 0 & delta hat(P) hat(M) - hat(M) delta hat(P) + delta hat(C) Pi^* - Pi delta hat(C)^* = 0}) itemizes into
$$ s \, ( \kappa \, j (\delta \theta) + \mu \, \delta b_0 1_{\mathbb{R}^3}) \, \psi + m_0 \, \mu \,  \delta k = 0,
$$
$$ (\kappa \, \delta v + \mu \, \delta \bar{b}) \times \psi = 0,
$$
$$ - s (\mu \, j (\delta \theta) + \kappa \, \omega^2 \delta b_0 1_{\mathbb{R}^3})  \psi - m_0 \, \omega^2  \kappa \, \delta k = 0
$$
\begin{equation}
- s (\mu \, \delta v + \kappa \, \omega^2 \delta \bar{b}) \cdot \psi + m_0 \, \omega^2 (\mu \, \delta \xi - \kappa \, \delta \tau_0) = 0
\label{2nd eqn to find the isotropy group of hat(mu) for G_omega}
\end{equation}
Taking into account (\ref{delta b =}), the solution of this system of equation is of the form 
$$ \delta \tau = \alpha, \qquad 
   \delta k = 0, \qquad 
   \delta \xi = \frac{\kappa}{\mu} \, \alpha
$$
$$ \delta v = \beta \, \psi, \qquad
   \delta \theta = \mu \, \psi, \qquad
   \delta b_0 = 0, \qquad
   \delta \bar{b} = - \beta \, \frac{\mu}{\omega^2 \kappa} \, \psi
$$
where $\alpha, \beta$ are two scalars of arbitrary values. Then the dimension of the isotropy group of the momentum then the number of its independent invariant is 2. We can take as independent invariants the {\bf rest mass} and the \textbf{spin}
$$ m_0 = \sqrt{\hat{\Pi}^*\hat{\Pi}}, \qquad
   s = \frac{\sqrt{- (\mbox{pol}_{\hat{\mu}}  (\hat{J}_1))^* \mbox{pol}_{\hat{\mu}}  (\hat{J}_1)}}{\sqrt{\hat{\Pi}^*\hat{\Pi}} }
$$
It is worth to remark that the electric charge is the linear momentum along the fifth dimension but is not an invariant of the motion in the early Universe.

\section{Constructing the connection associated to the group $\hat{\mathbb{G}}_0$}


At every symmetry group $G$ is attached a family of covariant derivatives of which the connection matrices are defined on the $G$-principal bundle of the $G$-frames and valued in its Lie algebra $\mathfrak{g}$ (see for instance \cite{Kobayashi 1963}). We call them $G$-connections. For the group $\hat{\mathbb{G}}_1$, we recover the Levi-Civita free torsion connection used in Kaluza-Klein theory. A key-point of interest is \textbf{to determine the} $\hat{\mathbb{G}}_0$-\textbf{connections}, the difficulty being that the manifold $\hat{\mathcal{U}}$ is not Riemannian, and to use it to deduce the equation of motion of a particle in gravitation and electromagnetic fields.

\subsection{Zoom  out}

It is time  now to restore the primes canceled for convenience in Section \ref{Subsection Zoom in} after the zoom in. For instance, according to (\ref{hat(G)_0 =}) and (\ref{hat(Omega)_0 =}), the two tensors $\hat{\bm{G}_0}$ and $\hat{\bm{\Omega}}_0$ are represented at the small scale respectively by
\begin{equation}
  \hat{G}'_0 = \left[ {{\begin{array}{cc}
       G'_0 \hfill &  0 \hfill \\
       0 \hfill &  0 \hfill \\
   \end{array} }} \right], \qquad 
 \hat{\Omega}'_0 =
    \left[ {{\begin{array}{cc}
       0 \hfill  \\
       1  \hfill  \\
   \end{array} }} \right]
\label{hat(G)'_0 = & hat(Omega)'_0 =}
\end{equation}
with the new notation for the matrix (\ref{G =})
\begin{equation}
 G'_0 = 
\left[ {{\begin{array}{cc}
       1 \hfill &  0 \hfill \\
       0 \hfill &  - 1_{\mathbb{R}^3} \hfill \\
   \end{array} }} \right]
\label{G'_0 =}
\end{equation}

At first glance, we could think to come back to the macroscopic scale by zooming out with the inverse transformation law of (\ref{zoom in})
$$ \hat{V} = \hat{P}_\omega \hat{V}'
$$
where $\hat{P}_\omega$ is given by (\ref{hat(P)_omega =}), that leads to the singularity $\hat{V}^5 = 0$ when $\omega \rightarrow 0$.

\subsection{Hypotheses on the pullback connection}
\label{SubSection Hypotheses on the pullback connection}

In order to avert the singularity resulting from the zoom out without to lose the relevant information in the fifth dimension, we propose to pullback over the space-time $\mathcal{U}$ the $\hat{\mathbb{G}}_0$-connection $\hat{\nabla}$ on the tangent bundle $T \hat{\mathcal{U}}$. The number of $\hat{\mathbb{G}}_0$-connections is infinite. The purpose here is not to study their properties in generic terms and to classify them. We are interested only by those which satisfy certain physical requirements on the ground of hypotheses considered below, {\bf (H1)} to {\bf (H4)} on the underlying geometric structure of the Universe and {\bf (H5)} on the motion of a charged elementary particle in both the gravitation and electromagnetic fields.

\begin{itemize}
    \item [{\bf (H1)}] \textbf{the space-time} $\mathcal{U}$ \textbf{is a Riemannian manifold}. The metric $\bm{G}$ is represented in an orthonormal basis $(\bm{e}'_1, \cdots, \bm{e}'_4)$ of $T_{\bm{X}} \mathcal{U}$ by the matrix $G'_0$ of (\ref{G'_0 =}) of the Minkowski metric.
    \item [{\bf (H2)}]$\pi : \hat{\mathcal{U}} \rightarrow \mathcal{U}$ \textbf{is a principal} $U(1)$\textbf{-bundle} of fibers isomorphic to the unitary group $U(1) = \left\lbrace z = e^{i\, t}, 0 \leq t < 2 \, \pi \right\rbrace$. The vector flow $t \mapsto \hat{\bm{X}} = e^{i\, t} \cdot \hat{\bm{X}}_0 $ is generated by the vector field $\hat{\bm{X}} \mapsto \hat{\bm{\Omega}}_0 (\hat{\bm{X}})$. 
    Let $\hat{\bm{X}}$ a point of the fiber of $\hat{\mathcal{U}}$ over $\bm{X}$. We say that a basis $(\hat{\bm{e}}_1, \cdot,\hat{\bm{e}}_5) $ of $T_{\hat{\bm{X}}} \hat{\mathcal{U}} $ is a \textbf{fibered basis} if the set of $\bm{e}_i = (T \pi) \hat{\bm{e}}_i$ for $1\leq i \leq 4$ is a basis of $T_{\bm{X}} \mathcal{U}$ and $\hat{\bm{e}}_5$ is tangent to the fiber ($(T \pi)  \hat{\bm{e}}_5 = \bm{0}$).
    
    Let us consider the fibered basis $(\hat{\bm{e}}'_\alpha) = (\hat{\bm{e}}'_1, \cdots, \hat{\bm{e}}'_4, \bm{\hat{\Omega}}_0)$ of $T_{\hat{\bm{X}}} \hat{\mathcal{U}}$ such as $(\bm{e}'_1, \cdots, \bm{e}'_4)$ is an orthonormal basis as for {\bf (H1)}. The pullback of the metric
    $$   \hat{\bm{G}}_0 = \pi^* \bm{G}
    $$
     is a field of symmetric 2-covariant tensors (but not a metric!). $(\hat{\bm{e}}'_\alpha)$ is a $\hat{\mathbb{G}}_0$-basis, \textit{i.e.} a basis in which $\hat{\bm{G}_0}$ and $\hat{\bm{\Omega}}_0$ are represented by (\ref{hat(G)'_0 = & hat(Omega)'_0 =}).
    \item[{\bf (H3)}] $\hat{\nabla}$ \textbf{is a} $\hat{\mathbb{G}}_0$-\textbf{connection}. Let $\hat{\pi} : \mathcal{F} (\hat{\mathcal{U}}) \rightarrow \hat{\mathcal{U}}$  be the frame bundle of $\hat{\mathcal{U}}$, {\it i.e.} the bundle  whose fiber  over $\bm{\hat{X}}$ is the set of all basis of $T_{\bm{\hat{X}}} \hat{\mathcal{U}}$. A moving frame is a smooth section $\bm{\hat{X}} \mapsto (\hat{\bm{e}}'_\alpha)$ of $\mathcal{F}$. It is called a coordinate frame (or natural frame) if locally the basis is associated to a coordinate system (or local chart) $(\hat{X}^\alpha)$ and if so it is denoted $(\hat{\bm{\partial}}_\alpha)$.
    The set $\mathcal{F}_{\hat{\bm{G}_0}}$ of $\hat{\mathbb{G}}_0$-bases is called $\hat{\mathbb{G}}_0$\textbf{-structure}. It is a subbundle of $\mathcal{F} (\hat{\mathcal{U}})$ and a principal $\hat{\mathbb{G}}_0$-bundle that is endowed with an \textbf{Ehresmann connection}. 

    As we are concerned only by the set of basis changes, {\it i.e.} the Lie subgroup  of linear transformations $\hat{a}= (0, \hat{P})$ of $\hat{\bm{G}_0}$, the corresponding infinitesimal generators of $\hat{\mathfrak{g}}_0$ are of the form 
    $\delta \hat{a}= (0, \delta \hat{P})$. A $\hat{\mathbb{G}}_0$-connection $\hat{\nabla}$ is defined by the field of $\hat{\mathfrak{g}}_0$-valued connection 1-form $(\hat{\bm{e}}'_\alpha) \mapsto (0, \hat{\Gamma}')$ on $\mathcal{F}_{\hat{\bm{G}_0}}$ such that
    $$ \hat{\nabla} \hat{\bm{e}}'_\alpha = \hat{\Gamma}'^\beta_\alpha \hat{\bm{e}}'_\beta
    $$
    According to (\ref{hat(Z) in hat(g)_0 equivalent to}), the value of the connection 1-form is the $5 \times 5$ matrix 
    \begin{equation}
         \hat{\Gamma}' =
    \left[ \begin{array}{cc}
       \Gamma' \hfill &  0 \hfill \\
       \Gamma'^5 \hfill &  0 \hfill \\
   \end{array}  \right]
    \label{hat(Gamma)' = }
    \end{equation}
    where $\Gamma'$ is a $4 \times 4$ matrix of 1-forms, skew-adjoint with respect to the metric (\ref{G'_0 =}), $\Gamma'^5$ is a 4-row of 1-forms and the last column of $\hat{\Gamma}'$ vanishes.
    Because $\hat{G}'_0$ and $\hat{\Omega}'_0$ are constant in a moving $\hat{\mathbb{G}}_0$-frame $\bm{\hat{X}} \mapsto (\hat{\bm{e}}'_a)$, we have 
    $$ d\hat{G}'_0 = 0, \qquad
       d\hat{\Omega}'_0 = 0
    $$
    Otherwise, these conditions must be replaced by 
    $$\hat{\nabla} \hat{G}_0 = 0,\qquad
      \hat{\nabla} \hat{\Omega}_0 = 0
    $$
    the connection $\hat{\nabla}$ preserving the pullback $\hat{\bm{G}}_0$ of the space-time metric and the vector $\hat{\Omega}_0$
    $$ \hat{\nabla} \,  \hat{\bm{G}}_0 = \bm{0}, \qquad
        \hat{\nabla} \, \hat{\Omega}_0 = \bm{0}
    $$
    \item [{\bf (H4)}]  Let $f$ be a section of the principal $U(1)$-bundle $\hat{\mathcal{U}}$, then $\bm{G} = f^* \hat{\bm{G}}_0$. The pullback bundle $f^* \, T \hat{\mathcal{U}}$ is a bundle over $\mathcal{U}$ of which the fiber over $\bm{X} \in \mathcal{U}$ is 
    $(f^*\,  T \hat{\mathcal{U}})_{\bm{X}} = T_{f(\bm{X})} \hat{\mathcal{U}}$. For any section $\hat{\bm{W}}$ of $T\hat{\mathcal{U}}$, the pullback section is $f^* \hat{\bm{W}} = \hat{\bm{W}} \circ f$. \textbf{The space-time is endowed with  the pullback connection} $f^* \hat{\nabla}$, determined uniquely by 
    $$ (f^* \hat{\nabla})_{\bm{U}} (f^* \hat{\bm{W}}) = f^* (\hat{\nabla}_{(Tf)\bm{U}} \hat{\bm{W}})
    $$
    and \textbf{torsion free}, then for all smooth fields $\bm{X} \mapsto \bm{U}, \bm{X} \mapsto \bm{V}$   of tangent vectors to the space-time $\mathcal{U}$
    \begin{equation}
     \hat{\nabla}_{(Tf) \, \bm{U}} (Tf) \bm{V}
    - \hat{\nabla}_{(Tf) \, \bm{V}} (Tf) \bm{U}
    - (Tf) \,\lbrack \bm{U}, \bm{V}\rbrack 
    = \bm{0}
    \label{torsion free condition}
    \end{equation}
    The section of the fiber bundle $\pi : \hat{\mathcal{U}} \rightarrow \mathcal{U}$ is an embedding  of $\mathcal{U}$ into $\hat{\mathcal{U}}$. The subspace $H_{f(\bm{X})} = (T f) T_{\bm{X}}\mathcal{U} $ of $T_{f(\bm{X})}\hat{\mathcal{U}}$ is isomorphic to $T_{\bm{X}}\mathcal{U}$ then of dimension 4 and the tangent space to the fiber $\hat{\mathcal{U}}_{\bm{X}}$ at $f(\bm{X})$ is the kernel of $T \pi$. Then $ T_{f(\bm{X})}\hat{\mathcal{U}} = H_{f(\bm{X})} \oplus T_{f(\bm{X})} \hat{\mathcal{U}}_{\bm{X}} $.
      A \textbf{basis adapted to the section} $f$ is a fibered basis $(\hat{\bm{e}}_1, \cdot,\hat{\bm{e}}_5) $ such that   $(\hat{\bm{e}}_1, \cdot,\hat{\bm{e}}_4) $ is a basis of $H_{f(\bm{X})}$. A way to build such a basis is to choose a non null vector $\hat{\bm{e}}_5$ tangent to the fiber $\hat{\mathcal{U}}_{\bm{X}}$ and a basis $(\bm{e}_1, \cdot,\bm{e}_4) $ of $T_{\bm{X}}\mathcal{U}$. Then, with the convention that Greek indices run from 1 to 5 and Latin ones run from 1 to 4, $(\hat{\bm{e}}_1, \cdot,\hat{\bm{e}}_4, \hat{\bm{e}}_5) = ((T f) \bm{e}_1, \cdot,(T f) \bm{e}_4, \hat{\bm{e}}_5) $ is a basis adapted to the section $f$ and
      \begin{equation}
           \bm{U} =  U^i \bm{e}_i \quad \Leftrightarrow \quad (T f) \bm{U} =  U^i \hat{\bm{e}}_i
      \label{vec(U) = sum & (Tf) vec(U) = sum}
      \end{equation}
      In terms of the dual bases, one has
      \begin{equation}
           \bm{e}^i (\bm{U}) 
        = \hat{\bm{e}}^i ((T f) \bm{U}) = U^i, \quad
          \hat{\bm{e}}^5 ((T f) \bm{U}) = 0
      \label{e^i (vec(U)) =}
      \end{equation}
      Using Christoffel's symbols, the elements of the connection 1-form in this basis are $\hat{\Gamma}^\beta_\alpha = \hat{\Gamma}^\beta_{\rho\alpha} \bm{e}^\rho$.
      In  particular, considering pairs of basis vectors $\bm{e}_i, \bm{e}_j$, the free torsion condition (\ref{torsion free condition}) reads 
      \begin{equation}
     \Gamma^\mu_{ij} 
         - \Gamma^\mu_{ji} - \delta^\mu_k c^k_{ij} 
         = 0, \qquad
     \label{Gamma^mu_(ij) - Gamma^mu_(ji) - delta^mu_k c^k_(ij) = 0}
      \end{equation}
    where $c^k_{ij}$ are the structure coefficient of the moving frame.
    \item [{\bf (H5)}] At this point, it is worth to remark that (\ref{Pi = P (Pi' - q' b)}) and (\ref{q = q'}) show that the 5-row 
    \begin{equation}
    	 \hat{\Pi} = \lbrack \Pi^*, q\rbrack
    \label{hat(Pi) = ( Pi^* , q)}
    \end{equation}
    represents a \textbf{linear form} $\hat{\bm{\Pi}}$ that we call \textbf{linear 5-momentum}.
    Let $s \mapsto \bm{X} (s)$ be the trajectory of a particle in the space-time, parameterized by the arc length $s$ (for the metric $\bm{G}$) and 
    \begin{equation}
         \bm{U} = \frac{d\bm{X}}{ds} 
    \label{vec(U) = dX / ds}
    \end{equation}
    is the unit tangent vector such that $\bm{U}^* \bm{U} = 1$.  
    We claim that
    \textbf{the motion of a charged particle and the evolution of its charge} $s \mapsto q (s)$ \textbf{are such that its linear 5-momentum is parallel-transported}
    \begin{equation}
         (f^* \hat{\nabla})_{\bm{U}} (f^* \hat{\bm{\Pi}}) = \bm{0}
    \label{parallel-transporded linear 5-momentum}
    \end{equation}
\end{itemize}

\subsection{Explicit form of the connection and the equation of motion}
\label{SubSection Explicit form of the connection and the equation of motion}

Our goal now is to determine explicitly the connection satisfying the hypotheses {\bf (H3)} to {\bf (H5)}. As we know, the $\hat{\mathbb{G}}_0$-structure $\mathcal{F}_{\hat{\bm{G}_0}}$ is in general non integrable. Then, instead of the moving frame $(\hat{\bm{e}}'_1, \cdots, \hat{\bm{e}}'_5)$ of {\bf (H3)}, section of $\mathcal{F}_{\hat{\bm{G}_0}}$,
we would like to work with a coordinate frame 
$(\bm{\partial}_1, \cdots, \bm{\partial}_4)$  
associated to a coordinate system $(X^1, \cdots, X^4)$ of the space-time $\mathcal{U}$
and the corresponding  fibered frame
$(\hat{\bm{e}}_1, \cdots, \hat{\bm{e}}_5)$ adapted to the section $f$, such as $   \hat{\bm{e}}_i = (T f) \bm{\partial}_i$ and obtained by the change of basis 
\begin{equation}
     \hat{\bm{e}}_\alpha 
   = \hat{P}^\beta_\alpha \hat{\bm{e}}'_\beta
\label{change of basis}
\end{equation}
where $\hat{P}^\beta_\alpha$ is the element at the intersection of the $\beta$-th row and the $\alpha$-th column of the transformation matrix partitioned into blocks as follows
\begin{equation}
     \hat{P} =
    \left[ \begin{array}{cc}
       P \hfill &  0 \hfill \\
       - 2 \, A^* \hfill &  1 \hfill \\
   \end{array}  \right]
\label{hat(P) = ((P , O), (A^* , 1))}
\end{equation}
where $P$ is a $4 \times 4$ matrix, $A$ is a 4-column. For $\hat{P}$  being regular, $P$ must be regular.
As the moving frames occurring in (\ref{change of basis}) are pulled back on the space-time $\mathcal{U}$, $P$ and $A$ are functions of the coordinates $X = (X^1, \cdots, X^4)$.
Owing to (\ref{hat(G)'_0 = & hat(Omega)'_0 =}), the pullback $\hat{\bm{G}} = \pi^* \bm{G}$ of the metric is represented in the new frame by 
$$ \hat{G} = \hat{P}^T \hat{G}'_0 \, \hat{P}
    = \left[ {{\begin{array}{cc}
       P^T G'_0 P \hfill &  0 \hfill \\
       0 \hfill & 0  \hfill \\
   \end{array} }} \right]
$$
where the $4 \times 4$ matrix
\begin{equation}
     G = P^T G'_0 P
\label{G = P^T G'_0 P}
\end{equation}
is regular. As $G'_0$ is constant and $P$ is a function of the coordinates $X$ on the time-space $\mathcal{U}$, $G$ is also a function of $X$ (but does not depends on $X^5$). 
In the new frame, the connection matrix is given by
$$ \hat{\Gamma} 
  = \hat{P}^{-1} ( \hat{\Gamma}'\, \hat{P} + d\hat{P})
$$
which, taking into account (\ref{hat(Gamma)' = }), leads to
\begin{equation}
               \hat{\Gamma} =
    \left[ \begin{array}{cc}
       \Gamma \hfill &  0 \\
       \Gamma^5 \hfill & 0 \hfill \\
   \end{array}  \right]
\label{hat(Gamma) = ((Gamma 0 (Gamma^5 0))}
\end{equation}
where
\begin{equation}
     \Gamma = P^{-1} (\Gamma'\, P + dP), \qquad 
   \Gamma_5 = \Gamma'^5 \, P 
             - 2 \, \lbrack d A^* 
             - A^* P^{-1} (\Gamma'\, P + dP) \rbrack
\label{Gamma = & Gamma_5 = & Gamma^5_5 =}
\end{equation}
Owing to (\ref{e^i (vec(U)) =}), the column representing $(T f) \bm{U}$ in the basis $(\hat{\bm{e}}'_i)$ is
$$ U' = P \, U 
      = P_i  U^i
$$
where  $P_i$ is the $i$-th column of $P$. Then, taking into account (\ref{Gamma = & Gamma_5 = & Gamma^5_5 =}) and the linear dependence of $\Gamma'$ with respect to $\hat{U}'$
$$ \Gamma (U) = P^{-1} (\Gamma' (U')\, P + dP)
            = \Gamma_i U^i
$$
where occurs the $4 \times 4$ matrix
$$ \Gamma_i = P^{-1} \left\lbrack \Gamma' (P_i)\, P + \partial_i P \right\rbrack
$$
where $\partial_i$ is the partial derivative with respect to $X^i$.
Owing to (\ref{G = P^T G'_0 P}), we have
$$ G \, \Gamma_i = P^T G'_0  \Gamma' (P_i)\, P 
                 + P^T G'_0 \partial_i P
$$
As $\Gamma'$ is skew-adjoint with respect to the metric (\ref{G'_0 =}), the first term of the right hand member is skew-symmetric. Then
$$  G \, \Gamma_i + (G \, \Gamma_i)^T
 = P^T G'_0 \partial_i P  
   + \partial_i P^T\, G'_0 P 
 = \partial_i G
$$
or
$$ \partial_i G
   - G \, \Gamma_i - (\Gamma_i)^T G = 0
$$
and with index notation
\begin{equation}
      \nabla_i G_{jk} 
    = \partial_i G_{jk} 
    - G_{jm} \Gamma^m_{ik} 
    - G_{mk} \Gamma^m_{ij} = 0
\label{nabla_i G_(jk) = 0}
\end{equation}
Besides, as we are working in a coordinate frame ($c^k_{ij} = 0$), the torsion free condition (\ref{Gamma^mu_(ij) - Gamma^mu_(ji) - delta^mu_k c^k_(ij) = 0}) gives  
\begin{equation}
     \Gamma^k_{ij} =  \Gamma^k_{ji}
\label{Gamma^k_(ij) =  Gamma^k_(ji)}
\end{equation}
Using the fundamental theorem of Riemannian geometry, both previous relations show that $\nabla$ is \textbf{Levi-Civita connection} 
\begin{equation}
     \Gamma^k_{ij} = \frac{1}{2} \, G^{mr}  \left\lbrack   \partial_j G_{ir}
+  \partial_i G_{jr}
-  \partial_r G_{ij}   \right\rbrack
\label{Levi-Civita connection}
\end{equation}
Then, owing to (\ref{Gamma = & Gamma_5 = & Gamma^5_5 =}), 
\begin{equation}
          \hat{\Gamma} (U) =
    \left[ \begin{array}{cc}
       \Gamma (U) \hfill & 0 \hfill \\
    \Gamma^5 (U)  \hfill & 0 \hfill \\
   \end{array}  \right]
\label{hat(Gamma) (U) =}
\end{equation}
where
\begin{equation}
    \Gamma^5 (U) = \Gamma'^5 (P \, U) \, P - 2 \, \nabla_U A^*
\label{Gamma^5 (U) = }
\end{equation}
In the moving frame $(\hat{\bm{e}}'_1, \cdots, \hat{\bm{e}}'_5)$ of {\bf (H3)}, section of $\mathcal{F}_{\hat{\bm{G}_0}}$, the linear 5-momentum $\hat{\bm{\Pi}}$ is represented by the 5-row
$\hat{\Pi}' = \lbrack \Pi'^*, q'\rbrack
$
where, restoring the primes, (\ref{Pi = m_0 with U = dX/ds such that U^* U = 1}) reads
$$      \Pi' = m_0 \, U' \quad 
    \mbox{with} \quad U' = \frac{dX}{ds} \quad
    \mbox{such that} \quad
    U'^* U' = 1
$$
In the fibered frame
$(\hat{\bm{e}}_1, \cdots, \hat{\bm{e}}_5)$ adapted to the section $f$, the linear 5-momentum is represented by $\hat{\Pi} = \lbrack \Pi^*, q\rbrack$ given by the transformation law of linear forms
$$ \hat{\Pi}  = \hat{\Pi}' \, \hat{P}
$$
with (\ref{hat(P) = ((P , O), (A^* , 1))}), that gives
$$ \Pi^* = m_0 \, U'^* P - 2 \, q' A^*, \qquad 
    q = q'
$$
hence the former relation can be recast as
\begin{equation}
     \Pi^* = m_0 \, U^* - 2 \,  q \, A^*
\label{Pi^* = m_0 U^* + q A^*}
\end{equation}
In the considered bases, the covariant derivative reads
$$ \hat{\nabla}_U \hat{\Pi} 
   = \dot{\hat{\Pi}} - \hat{\Pi} \, \hat{\Gamma} (U)  
$$
where $\dot{\hat{\Pi}}$  is the derivative of $\hat{\Pi}$ with respect to the arc length $s$ that is the proper time of the particle because $c = 1$.

According to hypothesis {\bf (H5)}, (\ref{parallel-transporded linear 5-momentum}), (\ref{hat(Gamma) (U) =}) and (\ref{Gamma^5 (U) = }) lead to 
\begin{equation}
   \nabla_U \Pi^* 
   - q \, (\Gamma'^5 (P \, U) \, P - 2 \, \nabla_U A^*) = 0 
\label{nabla Pi = 0}
\end{equation}
\begin{equation}
   \dot{q} = 0 
\label{nabla q = 0}
\end{equation}
As expected, according to the experience, \textbf{the electric charge} $q$ \textbf{is an integral of the motion}. Then, using (\ref{Pi^* = m_0 U^* + q A^*}), (\ref{nabla Pi = 0}) reads
\begin{equation}
    \forall m_0 \neq 0, \;\; \forall q, \;\;
    \forall U \; \mbox{such that} \; U^* U = 1,\qquad
   \nabla_U (m_0 \, U^*) 
   - q \, \Gamma'^5 (P \, U) \, P  = 0 
\label{nabla Pi = 0 simplified}
\end{equation}
In particular, if $q = 0$, we have
$$  \forall m_0 \neq 0,  \;\;
    \forall U \; \mbox{such that} \; U^* U = 1,\qquad
    \dot{m}_0 \, U^* + m_0 \, \nabla_U U^*  = 0
$$
Right multiplying by $U$ gives
$$  \forall m_0 \neq 0,  \;\;
    \forall U \; \mbox{such that} \; U^* U = 1,\qquad
    \dot{m}_0  + m_0 \, (\nabla_U U^*) \, U  = 0
$$
but as $U^* U = 1$ and $\nabla_U G = 0$, we have
$$ (\nabla_U U)^* U = (\nabla_U U^*)\, U
                = - U^* (\nabla_U U)
                = - (\nabla_U U)^* U
$$
then $(\nabla_U U)^* U$ vanishes and, consequently 
\begin{equation}
    (\nabla_U U^*) \, U = 0
\label{(nabla_U U^*) U = 0}
\end{equation}
that entails
$$ \dot{m}_0 = 0
$$
As expected, according to the experience, \textbf{the rest mass} $m_0$ \textbf{is an integral of the motion}\footnote{However, it is possible to extend the formalism to a particle with time-varying mass by introducing a thrust force as in \cite{AffineMechBook}}.
Besides, (\ref{nabla Pi = 0 simplified}) is reduced to
\begin{equation}
    \forall m_0 \neq 0, \;\; \forall q, \;\;
    \forall U \; \mbox{such that} \; U^* U = 1,\qquad
   m_0 \,\nabla_U  U^* 
   - q \, \Gamma'^5 (P \, U) \, P  = 0 
\label{nabla Pi = 0 simplified BIS}
\end{equation}
Owing to (\ref{(nabla_U U^*) U = 0}), right multiplying by $U$ gives  for $q \neq 0$
$$ \Gamma'^5 (P \, U) \, P \, U = 0
$$
The bilinear form 
$$ \varphi (U, V) = \Gamma'^5 (P \, U) \, P \, V 
$$
is such that $\varphi (U, U) = 0$ for all $U$ then it is skew-symmetric. There exists a $4 \times 4$ matrix $F$ such as
\begin{equation}
     F^T = - F \qquad
   \varphi (U, V) = U^T F \, V, \qquad
   \Gamma'^5 (P \, U) \, P = U^T F
\label{F^T = - F & phi (U,V) = U^T F V & Gamma'^5 (P U) P = U^T F}
\end{equation}
Moreover, (\ref{Gamma^5 (U) = }) becomes
$$   \Gamma^5 (U) = U^T F - 2 \, \nabla_U A^*
$$
The $j$-th component of this 4-row is
$$     \Gamma^5_j (U) = U^k F_{kj} - 2 \, \nabla_U A_j
$$
that gives rise, $(e_i)$ being the canonical basis of $\mathbb{R}^4$, to the corresponding Christoffel's symbols
\begin{equation}
      \Gamma^5_{ij} = \Gamma^5_j (e_i) 
     = \delta^k_i F_{kj} - 2 \, \nabla_i A_j
     = F_{ij} - 2 \, \nabla_i A_j
\label{Gamma^5_(ij) = F_(ij) - 2 nabla_i A_j}
\end{equation}
The torsion free condition (\ref{Gamma^mu_(ij) - Gamma^mu_(ji) - delta^mu_k c^k_(ij) = 0}) gives  
\begin{equation}
     \Gamma^5_{ij} -  \Gamma^5_{ji} 
    = F_{ij} - F_{ji} - 2 \, ( \nabla_i A_j - \nabla_j A_i)
    = 2 \, \lbrack F_{ij}  - ( \nabla_i A_j - \nabla_j A_i) \rbrack = 0
\label{Gamma^5_(ij) - Gamma^5_(ji) =}
\end{equation}
where, because of the torsion free condition (\ref{Gamma^k_(ij) =  Gamma^k_(ji)})
$$ \nabla_i A_j - \nabla_j A_i 
   = \left( \partial_i A_j   - \Gamma^k_{ij} A_k \right)  
  - \left( \partial_j A_i  - \Gamma^k_{ji} A_k \right)  
  =  \left( \partial_i A_j -  \partial_j A_i   \right)  
  -  (\Gamma^k_{ji}- \Gamma^k_{ij}) \, A_k   
 =  \partial_i A_j -  \partial_j A_i  
$$
Then the skew-symmetric $4 \times 4$ matrix $F$ of elements 
\begin{equation}
     F_{ij} = \partial_i A_j -  \partial_j A_i  
\label{F_(ij) = partial A_j / partial X^i -  partial A_i / partial X^j }
\end{equation}
represents a skew-symmetric 2-covariant tensor (or 2-form) $\bm{F}$ and, owing to (\ref{F^T = - F & phi (U,V) = U^T F V & Gamma'^5 (P U) P = U^T F}), the equation (\ref{nabla Pi = 0 simplified BIS}) becomes
$$ m_0 \,\nabla_U  U^* = q \, U^T F   
$$
by transposition and left multiplication by $G^{-1}$, we obtain the \textbf{equation of motion}
\begin{equation}
    m_0 \,\nabla_U  U = - q \, \bar{F} \, U
\label{m_0 nabla_U U = (1/2) q F U}
\end{equation}
where the skew-adjoint matrix 
$$ \bar{F} = G^{-1} F
$$
represents a 1-covariant and 1-contravariant tensor $\bar{\bm{F}}$ of components $F^i_j$. 
The physical interpretation is that the vector $\bm{A}$ of which the adjoint $\bm{A}^*$ is represented by the row $A^*$ is the \textbf{electromagnetic} 4-\textbf{potential}, $\bm{F}$ (resp. $\bar{\bm{F}}$) is the 2-covariant (resp. 1 contravariant and 1-covariant) \textbf{electromagnetic field}
\footnote{with the conventions used in \S 15 of \cite{SSD, SSDEng}} 
 and 
$$ \bm{f} = - q \, \bar{\bm{F}} \cdot \bm{U}
$$
is the \textbf{Lorentz force}, represented by the right hand side of (\ref{m_0 nabla_U U = (1/2) q F U}). We recover the equation of motion of a charged particle in the gravitation and electromagnetic fields that fits the observations today. It deserves to remember that the equation (\ref{parallel-transporded linear 5-momentum}) itemizes into a group of two equations, the former one being the previous equation of motion and the later one (\ref{nabla q = 0}) being  the \textbf{conservation of the charge}
$$ \dot{q} = 0
$$
The condition (\ref{F_(ij) = partial A_j / partial X^i -  partial A_i / partial X^j }) means that the 2-form $\bm{F}$ is the exterior derivative of the 1-form $\bm{A}^*$.
The 4-potential $\bm{A}$ and its adjoint $\bm{A}^*$ being represented respectively by
$$ A = 
    \left[ \begin{array}{c}
       \phi \hfill  \\
       \mathsf{A} \hfill  \\
   \end{array}  \right], \qquad
   A^* = \lbrack \phi, - \mathsf{A}^T \rbrack
$$
the electromagnetic fields $\bm{F}$  and $\bar{\bm{F}}$ are represented respectively by
$$  F = 
    \left[ \begin{array}{cc}
       0 \hfill & - E^T \\
       E \hfill & - j(B) \\
   \end{array}  \right], \qquad
  \bar{F} = 
    \left[ \begin{array}{cc}
       0 \hfill & - E^T \\
       - E \hfill & j(B) \\
   \end{array}  \right]
$$
where
\begin{equation}
     E = - grad \, \phi - \partial_t \mathsf{A}, \qquad
   B = curl \, \mathsf{A}
\label{E & B in terms of potentials}
\end{equation}
where $\partial_t$ is the partial derivative to the time.
Owing to (\ref{U = dX / ds}), the Lorentz force $\bm{f}$ is represented by
$$ f = - q \, \bar{F} \, U =
   \gamma \, q \, 
   \left[ \begin{array}{c}
       E \cdot v \hfill  \\
       E + v \times B \hfill  \\
   \end{array}  \right]
$$

{\bf Remark 1.} The equation (\ref{m_0 nabla_U U = (1/2) q F U}) is the usual form of the equation of motion as found in the literature. It has been obtained by a suitable choice of adapted bases but it is not the most general form according to the covariant equation (\ref{parallel-transporded linear 5-momentum}). Of course, it is possible to make other choices but it is not necessarily interesting.

{\bf Remark 2.} It can be verified that, as expected, 
 the connection $\hat{\nabla}$ preserves in the adapted bases the pullback $\hat{\bm{G}}_0$ of the space-time metric and the vector $\hat{\bm{\Omega}}_0$
$$ \hat{\nabla} \,  \hat{G}_0 = 0, \qquad
        \hat{\nabla} \, \hat{\Omega}_0 
        = d\hat{\Omega}_0 + \hat{\Gamma} \, \hat{\Omega}_0  = 0
$$
For $\hat{\bm{G}}_0$, it results from (\ref{nabla_i G_(jk) = 0}) while for $\hat{\bm{\Omega}}_0$ it is a consequence of (\ref{hat(Gamma) = ((Gamma 0 (Gamma^5 0))}) and 
$$ \hat{\Omega}_0 =
    \left[ {{\begin{array}{cc}
       0 \hfill  \\
       1  \hfill  \\
   \end{array} }} \right]
$$
resulting from (\ref{hat(G)'_0 = & hat(Omega)'_0 =}) and (\ref{hat(P) = ((P , O), (A^* , 1))}).

{\bf Remark 3.} Finally, it is worth to remark that, although the choice (\ref{hat(P) = ((P , O), (A^* , 1))})  of $\hat{P}$  is not general, it is sufficient to guarantee the existence and uniqueness of the connection in terms of the fields  $G$ and  $A^*$  thanks to the free torsion condition. On the other hand, this choice depends on the section $f$. It is what we are going to discuss now.

\subsection{Gauge transformation}

 The coordinate basis $(\bm{\partial}_i)$ and $\hat{\bm{e}}_5$ being given, the adapted basis $(\hat{\bm{e}}_\alpha)$ of $\mathcal{U}$ to $f$ is uniquely defined by the section $f$. We would like to know to what extend the equation of motion does not depend on the choice of the section $f$. For this aim, we consider another section $f''$ and the corresponding adapted basis $(\hat{\bm{e}}''_\alpha)$. As $(T \pi) \hat{\bm{e}}_i = (T \pi) \hat{\bm{e}}''_i = \bm{\partial}_i$, the vector $\hat{\bm{e}}''_i - \hat{\bm{e}}_i $ is tangent to the fiber of $\mathcal{U}$ over $\bm{X}$ then 
 $$ \hat{\bm{e}}''_i - \hat{\bm{e}}_i = b_i \hat{\bm{e}}_5, \qquad
 \hat{\bm{e}}''_5 = \hat{\bm{e}}_5
 $$
 the change of basis from $(\hat{\bm{e}}_\alpha)$ to $(\hat{\bm{e}}''_\alpha)$ is given by the transformation matrix 
 $$ \hat{P}_{f \rightarrow f''} = 
    \left[ \begin{array}{cc}
       1_{\mathbb{R}^4} \hfill & 0 \hfill \\
       b^* \hfill & 1 \hfill \\
   \end{array}  \right]
 $$
  where $b^*$ is the 4-row of components $b_i$. 
The set of these matrices is an Abelian group $\hat{\mathbb{G}}_b$ for the matrix product. To determine whether the $\hat{\mathbb{G}}_b$-structure is integrable, we have to solve the PDE system
$$ \frac{\partial \hat{X}''}{\partial \hat{X}} = \hat{P}_{f \rightarrow f''}, \qquad 
    \hat{P}_{f \rightarrow f''} \in \hat{\mathbb{G}}_b
$$
that gives rise to
$$ \frac{\partial X''}{\partial X} = 1_{\mathbb{R}^4} ,\qquad 
   \frac{\partial X''}{\partial X^5} = 0, \qquad 
 \frac{\partial X''^5}{\partial X} = - b ,  \qquad
   \frac{\partial X''^5}{\partial X^5} = 1
$$
The two former equations lead to $X" = X + X_0$ 
where the constant $X_0$ may be discarded without inconvenience 
$$ X" = X 
$$
The last equation gives
$$ X"^5 = X^5  + 2 \, h (X)
$$
where $h$ is an arbitrary smooth function. 
Finally, the third equation provides the integrability condition 
$$  b^* = - 2 \, \frac{\partial h}{\partial X} 
$$
The linear form $\bm{\Phi}$ represented in the basis $(\hat{\bm{e}}_\alpha)$ by the 5-arrow
 $$ \Phi = \lbrack A^*, 1\rbrack
 $$
 is represented in the basis $(\hat{\bm{e}}''_\alpha)$ by the 5-arrow $\Phi'' = \Phi \, \hat{P}_{f \rightarrow f''} $, leading to the \textbf{gauge transformation} 
\begin{equation}
     A''^{*} = A^* + \frac{\partial h}{\partial X}
\label{A*'' = A* + partial h / partial X}
\end{equation}
 As the matrix (\ref{G = P^T G'_0 P}) does not depends on the fifth coordinate, $G$ is not modified by the coordinate change as well as, owing to (\ref{F_(ij) = partial A_j / partial X^i -  partial A_i / partial X^j }), $\bar{F}$  and $F = G^{-1} F$. Hence {\bf the choice of the section} $f$ {\bf does not modify the equation of motion} (\ref{m_0 nabla_U U = (1/2) q F U}).

\section{Extended variational relativity}

The goal of this section is to extend the general theorems of the relativity based on Einstein-Hilbert functional \cite{Hilbert 1915}. We use Palatini formalism, considering the connection as an unknown independent field \cite{Palatini 1919}.  

\subsection{Stationary action principle}

Inspired by the presentation given in \cite{GR}, we start from the following principles
\begin{itemize}
     \item [{\bf (P1)}] To every {\bf physical phenomenon} corresponds a field $\bm{z}$, represented in a local chart by a $n_{(z)}$-column $z$, and a field 
     $$ (z, \partial_i z, G_{ij}, A_i) \mapsto 
        L_{(z)} (z, \partial_i z, G_{ij}, A_i) \in \mathbb{R}
     $$
     called {\it presence of the phenomenon} in \cite{GR}.
     \item [{\bf (P2)}] For the Lagrangian
     $$ L = \sum_z L_{(z)}
     $$
     the \textbf{action}
     \begin{equation}
          S = \int_{\mathcal{D}} L \, vol
          = \int_{\mathcal{D}} L \, u \, \mbox{d}^4 X
          = \sum_z  S_{(z)} = \sum_z \int_{\mathcal{D}} L_{(z)} u \, \mbox{d}^4 X
     \label{action}
     \end{equation}
     is stationary for every variation of $G, A^*$ and the concomitant phenomena, null on the boundary of an open domain $\mathcal{D}$ of the space-time, $vol$ being the Riemannian volume 4-form of component $u$ in the local chart $(X^i)$
\end{itemize}
As the signature of Minkowski metric is $(+ - - -)$, $\det G < 0$, then
$$ u = \sqrt{- \det G }
$$
and the covariant divergence of any vector field $\bm{V}$ for Levi-Civita connection is the scalar field
\begin{equation}
     \mbox{div} \, \bm{V} 
     = \nabla_i V^i
     = \frac{1}{u} \, \partial_i (u \, V^i)
\label{covariant divergence of a vector}
\end{equation}
By differentiation of the presence, we have 
\begin{equation}
	 \delta L_{(z)} = p \, \delta z 
     + \sigma^j \delta ( \partial_j z) 
     + E^{km}_{(z)} \delta G_{km}
     + \tilde{T}^k_{(z)} \delta A_k
\label{delta L_(z) = p delta z + sigma^j delta (partial z) + E^jk_(z) delta _jk + tilde(T)^^k_(z) A_k}
\end{equation}
where $p, \sigma^j$ are $n_{(z)}$-rows,  $E^{km}_{(z)} = E^{mk}_{(z)}$ and $\tilde{T}^k_{(z)}$ are the components of a vector field $\tilde{\bm{T}}_{(z)}$. Taking into account
$$ \frac{\delta u}{u} = \frac{1}{2} \, G^{ij} \delta G_{ij}
   = - \frac{1}{2} \, G_{ij} \delta G^{ij}
$$
we have
$$ \delta (L_{(z)} u) =  \lbrack p \, \delta z 
     + \sigma^j \delta ( \partial_j z) 
     + \frac{1}{2} \, T^{km}_{(z)} \delta G_{km}
     + \tilde{T}^k_{(z)} \delta A_k 
     \rbrack \, u
$$
where 
\begin{equation}
	 T^{jk}_{(z)} = 2 \, E^{jk}_{(z)} + L_{(z)} G^{jk}
\label{T^jk_(z) = 2 E^jk_(z)  + L_(z) G^jk}
\end{equation}
are the components of a symmetric 2-contravariant tensor field  $\bm{T}_{(z)}$. Introducing the $n_{(z)}$-row of scalar field
$$ W_{(z)}  =  p - \frac{1}{u} \, \partial_j (u \, \sigma^j )
                =  p - \nabla_j \sigma^j
$$
the vector field $\bm{V}_{(z)} (\delta \bm{z})$ represented by the 4-column $V_{(z)} (\delta z)$ of components
$$ V^j_{(z)}  (\delta z)  = \sigma^j \delta z
$$
and owing to (\ref{covariant divergence of a vector}), 
\begin{equation}
     \delta (L_{(z)} u) =  \lbrack W_{(z)} \,  \delta z 
     + \mbox{div} \, (V_{(z)} (\delta z))
     + \frac{1}{2} \, T^{km}_{(z)} \delta G_{km}
     + \tilde{T}^k_{(z)} \delta A_k 
     \rbrack \, u
\label{delta (L_(z) u) = }
\end{equation}
or, the expression being independent of the choice of the local chart
\begin{equation} 
    \delta (L_{(z)} u) =  \lbrack W_{(z)} \,  \delta z 
     + \mbox{div} \, (\bm{V}_{(z)} (\delta \bm{z}))
     + \frac{1}{2} \, \bm{T}_{(z)} :  \delta \bm{G}
     +  \delta \bm{A}^* \cdot \tilde{\bm{T}}_{(z)}
     \rbrack \, u    
\label{delta (L_(z) u) = BIS}
\end{equation}
that allows to prove 
\begin{theorem}
The stationary action principle {\bf (P2)} is equivalent to 
\begin{itemize}
    \item[(i)] $ W_{(z)}  = 0 \,$ for every phenomenon
    \item[(ii)] $\bm{T} = \sum_z \bm{T}_{(z)} = \bm{0}$
    \item[(ii)] $\tilde{\bm{T}} = \sum_z \tilde{\bm{T}}_{(z)} = \bm{0}$
\end{itemize}
\label{thm stationary action principle}
\end{theorem}
{\bf Proof.} Indeed, combining (\ref{action}) and (\ref{delta (L_(z) u) = BIS}), the variation of the action reads
$$ \delta S = \int_{\mathcal{D}} \delta (L \, u) \, \mbox{d}^4 X
      = \sum_z \delta S_{(z)}
$$
with
\begin{equation}
 \delta S_{(z)} = \int_{\mathcal{D}} \delta (L_{(z)}  u) \, \mbox{d}^4 X
   = \int_{\mathcal{D}} \lbrack W_{(z)} \,  \delta z 
     + \mbox{div} \, (\bm{V}_{(z)} (\delta \bm{z}))
     + \frac{1}{2} \, \bm{T}_{(z)} :  \delta \bm{G}
     +  \delta \bm{A}^* \cdot \tilde{\bm{T}}_{(z)}
     \rbrack \, vol
\label{delta S_(z) =}
\end{equation}
where, denoting $\mathcal{L}_{\bm{V}_{(z)}(\delta \bm{z})} $ the Lie derivative along the vector field $\bm{V}_{(z)} (\delta \bm{z})$, one has
$$  \int_{\mathcal{D}}  \mbox{div} \, (\bm{V}_{(z)} (\delta \bm{z})) \, vol 
  = \int_{\mathcal{D}}  \mathcal{L}_{\bm{V}_{(z)} (\delta \bm{z})} \, vol 
$$
Denoting $d$ the exterior derivative of a differential form and $\iota_{\bm{V}}$ the interior product with the vector $\bm{V}$, owing to Cartan's formula and Stokes' theorem
$$  \int_{\mathcal{D}}  \mbox{div} \, (\bm{V}_{(z)}(\delta \bm{z})) \, vol 
  = \int_{\mathcal{D}} d \lbrack  \iota_{\bm{V}_{(z)} (\delta \bm{z})} vol \rbrack
  = \int_{\partial\mathcal{D}}   \iota_{\bm{V}_{(z)} (\delta \bm{z})} vol 
$$
Considering smooth functions $\delta \bm{z}$ on compact supports contained in $\mathcal{D}$, then null on its boundary, this integral vanishes and it remains
$$ \delta S_{(z)}  = \int_{\mathcal{D}} \lbrack W_{(z)} \,  \delta z 
     + \frac{1}{2} \, \bm{T}_{(z)} :  \delta \bm{G}
     +  \delta \bm{A}^* \cdot \tilde{\bm{T}}_{(z)}
     \rbrack \, vol
$$
and the stationarity of the action entails
$$ \delta S  = \int_{\mathcal{D}} \lbrack \sum_z W_{(z)} \,  \delta z 
     + \frac{1}{2} \, \bm{T} :  \delta \bm{G}
     +  \delta \bm{A}^* \cdot \tilde{\bm{T}}
     \rbrack \, vol = 0
$$
Considering the only non vanishing variation of a particular phenomenon $z$ and applying the fundamental lemma of the calculus of variations, we prove (i). Likewise, considering the non vanishing compactly supported smooth variations of $\bm{G}$ (resp. $\bm{A}^*$), we prove (ii) (resp. (iii)). $\blacksquare$

\vspace{0.2cm}

\textbf{Remark.} It is worth to remark  that the number of field equations \textit{(i)} and \textit{(ii)}, fourteen, is the same as  the number of unknowns (independent components of $G$ and $A^*$), that avoids being caught in the dilemma of classical Kaluza-Klein theory: add the dilaton but one condition has no physical interpretation, or set the dilaton to one to recover Maxwell equations but this condition is violated. 

\vspace{0.2cm}

The principle of stationary action can be used in different ways. For instance, it is worth to notice that the equation (i)
\begin{equation}
      p - \nabla_j \sigma^j = 0
\label{p - nabla_j sigma^j = 0}
\end{equation}
can be obtained considering only the stationarity of the corresponding part $S_{(z)}$ of the action. It is this way that we are going to follow from now on. 
\begin{corollary}  
$\mbox{div} \, \tilde{\bm{T}}_{(z)} = 0 \, $ for every phenomenon
\label{Corollary}
\end{corollary}
{\bf Proof.} According to the gauge transformation (\ref{A*'' = A* + partial h / partial X}), we perform the substitution $A_i \rightarrow A_i + \partial_i h$ in the Lagrangian 
$L_{(z)}$ of the action $S_{(z)}$. Applying the equation (\ref{p - nabla_j sigma^j = 0}) to the phenomenon $h$ (then $p =0$ and $\sigma_j = \tilde{T}^j_{(z)}$) gives
\begin{equation}
     \nabla_j \tilde{T}^j_{(z)} = 0
\label{nabla_j tilde(T)^j_(z) = 0}
\end{equation}
that achieves the proof. $\blacksquare$

\subsection{Laws of conservation}

\begin{theorem}
The equations $ W_{(z)}  = 0 \,$ 
and $\mbox{div} \, \tilde{\bm{T}}_{(z)} = 0 \, $ 
entail the law of conservation 
\begin{equation}
 \mbox{div} \,  \bar{\bm{T}}_{(z)} + \tilde{\bm{T}}_{(z)} \cdot \bm{F} = \bm{0}
\label{law of conservation}   
\end{equation}
of the 1-contravariant and 1-covariant tensor $\bar{\bm{T}}_{(z)}$ of components $(T_{(z)})^i_j = (T_{(z)})^{ik} G_{kj}$.
\end{theorem}
{\bf Proof.} A smooth vector field $\delta \bm{X}$ of compact support contained in $\mathcal{D}$ is represented in a local chart by a 4-column $\delta X$ and leads to the variations $\mathcal{L}_{\delta X} \,  z, \mathcal{L}_{\delta X} \, G$ and $\mathcal{L}_{\delta X} \, A^*$.
Owing to (\ref{delta (L_(z) u) = }), the corresponding variation of the Lagrangian is given by
$$      \delta (L_{(z)} u) =  \lbrack W_{(z)} \, \mathcal{L}_{\delta X} \,  z  
     + \mbox{div} \, (V_{(z)} (\delta z))
     + \frac{1}{2} \, T^{km}_{(z)} (\mathcal{L}_{\delta X} \,  G)_{km}
     + \tilde{T}^j_{(z)} (\mathcal{L}_{\delta X} \,  A^*)_j 
     \rbrack \, u
$$
where, owing to (\ref{Levi-Civita connection})
$$ (\mathcal{L}_{\delta X} \,  G)_{km} 
   = \delta X^r \partial_r G_{km}
   + G_{kr} \partial_m \delta X^r
   + G_{rm} \partial_k \delta X^r
   = \nabla_k \delta X_m + \nabla_m \delta X_k
$$
Then, taking into account the symmetry of the tensor $\bm{T}_{(z)}$
$$ \frac{1}{2} \, T^{km}_{(z)} (\mathcal{L}_{\delta X} \,  G)_{km}
    =  T^{km}_{(z)} \, \nabla_k \delta X_m
    =  \nabla_k ( T^{km}_{(z)} \, \delta X_m) - (\nabla_k  T^{km}_{(z)} ) \, \delta X_m
$$
$$ \frac{1}{2} \, T^{km}_{(z)} (\mathcal{L}_{\delta X} \,  G)_{km} 
   =  \nabla_k ( (T_{(z)})^k_m \, \delta X^m) - (\nabla_k  (T_{(z)})^k_m ) \, \delta X^m
$$
or, the expression being independent of the choice of the local chart
\begin{equation}
     \frac{1}{2} \, \bm{T}_{(z)} :  \mathcal{L}_{\delta \bm{X}} \bm{G}
    =  \mbox{div} \, (\bar{\bm{T}}_{(z)} \cdot \delta \bm{X})
       - (\mbox{div} \,  \bar{\bm{T}}_{(z)}) \cdot \delta \bm{X}
\label{1/2 T_(z) : L_vec(delta X) G =}
\end{equation}

Besides, one has
$$ (\mathcal{L}_{\delta X} \,  A^*)_j 
   = \delta X^r \partial_r A_j
   + A_r \partial_j \delta X^r
$$
then, integrating by part
$$ u \,\tilde{T}^j_{(z)} (\mathcal{L}_{\delta X} \,  A^*)_j
    = \partial_j (u \,\tilde{T}^j_{(z)} A_r \delta X^r)
     - \lbrack \partial_j (u \,\tilde{T}^j_{(z)} A_r)
               - u \,\tilde{T}^j_{(z)} \partial_r A_j \rbrack \, \delta X^r
$$
Taking into account (\ref{F_(ij) = partial A_j / partial X^i -  partial A_i / partial X^j })
$$ u \,\tilde{T}^j_{(z)} (\mathcal{L}_{\delta X} \,  A^*)_j
    = \partial_j (u \,\tilde{T}^j_{(z)} A_r \delta X^r)
     - \lbrack \partial_j (u \,\tilde{T}^j_{(z)}) A_r
               + u \,\tilde{T}^j_{(z)} F_{jr})\rbrack \, \delta X^r
$$
and, owing to (\ref{covariant divergence of a vector}) and (\ref{nabla_j tilde(T)^j_(z) = 0}), we obtain
$$ u \,\tilde{T}^j_{(z)} (\mathcal{L}_{\delta X} \,  A^*)_j
    = \lbrack \nabla_j (\tilde{T}^j_{(z)} A_r \delta X^r)
     -  \tilde{T}^j_{(z)} F_{jr} \, \delta X^r
     \rbrack \, u
$$
or, the expression being independent of the choice of the local chart
\begin{equation}
 ( \mathcal{L}_{\delta \bm{X}} \bm{A}^*) \cdot \tilde{\bm{T}}_{(z)}
   = \mbox{div} \, ( (\bm{A}^* \cdot \delta \bm{X}) \,\tilde{\bm{T}}_{(z)})
   -    \tilde{\bm{T}}_{(z)} \cdot \bm{F} \cdot \delta \bm{X}
\label{(L_vec(delta X) A^*) cdot vec(tilde(T))_(z) = }
\end{equation}
Introducing into (\ref{delta S_(z) =}), with the variations given by Lie derivatives, the expression (\ref{1/2 T_(z) : L_vec(delta X) G =}) and (\ref{(L_vec(delta X) A^*) cdot vec(tilde(T))_(z) = }), and owing to Theorem \ref{thm stationary action principle} (i), we have
$$  \int_{\mathcal{D}} \mbox{div} \, \lbrack 
      \bm{V}_{(z)} \, (\mathcal{L}_{\delta \bm{X}} \bm{z})
        + \bar{\bm{T}}_{(z)} \cdot \delta \bm{X}
        + (\bm{A}^* \cdot \delta \bm{X}) \,\tilde{\bm{T}}_{(z)}
     \rbrack \, vol
- \int_{\mathcal{D}} \lbrack 
        \mbox{div} \,  \bar{\bm{T}}_{(z)}) 
        + \tilde{\bm{T}}_{(z)} \cdot \bm{F} 
     \rbrack  \cdot \delta \bm{X}
     \, vol = 0
$$
Considering smooth variations of compact support contained in $\mathcal{D}$ and reasoning as in the proof of Theorem \ref{thm stationary action principle}, the first integral vanishes and, applying  the fundamental lemma of the calculus of variations, we obtain (\ref{law of conservation}). $\blacksquare$

Our aim now is to apply these general principle by constructing Lagrangians for three phenomena, the matter, the gravitation and the electromagnetism.

\subsection{Laws of conservation of the matter}

\vspace{0.3cm}

\textbf{Matter manifold}

\vspace{0.3cm}

In \S 36 of \cite{GR} and \S 11.2 of \cite{Soper 1976}, the motion of a continuum is described by a line bundle $\pi_A: \mathcal{U}\rightarrow\mathcal{M}$ where $\mathcal{M}$ is a Riemannian manifold of dimension 3 representing the matter and called the \textbf{matter manifold}. Its metric is denoted $\bm{G}_m$. Each fiber is the trajectory of a \textbf{material particle} $\bm{a} = \pi_A (\bm{X})$. In a local chart of $\mathcal{M}$, it is represented by its \textbf{Lagrange coordinates} $ a \in \mathbb{R}^3$. As $a$ is an invariant of the motion, the material derivative vanishes:
$$   \frac{d \bm{a}}{ds} = (T \pi_A)  \cdot \bm{U} = \bm{0}
$$
where $\bm{U}$ is tangent to the trajectory and defined by (\ref{vec(U) = dX / ds}). In local charts of $\mathcal{U}$ and $\mathcal{M}$, 
$$ \frac{d a}{ds} 
     = \frac{\partial a}{\partial X} \, \frac{d X}{ds} 
       = M \, U = 0
$$
where, because the fiber is a line, $M$ is a $3 \times 4$ matrix of full rank and $U$ is given by (\ref{U = dX / ds}).

To extend this formalism to the 5D universe $\hat{\mathcal{U}}$, we consider a line bundle $\hat{\pi}_A: \hat{\mathcal{U}}\rightarrow \hat{\mathcal{M}}$ where $\hat{\mathcal{M}}$ is a manifold of dimension 4 representing the matter in this universe. By analogy with the hypothesis {\bf (H2)} of Section \ref{SubSection Hypotheses on the pullback connection}, we suppose that $\pi_M: \hat{\mathcal{M}}\rightarrow \mathcal{M}$ is a principal $U(1)$-bundle. Bearing in mind that we have to perform a pullback, we assume that $\pi_A \circ \pi = \pi_M \circ \hat{\pi}_A$, then the following diagram is commutative
\[
\begin{CD}
     \hat{\mathcal{U}}  @>{\hat{\pi}_A}>> \hat{\mathcal{M}} \\
    @V{\pi}VV                             @VV{\pi_M}V \\
    \mathcal{U}        @>>{\pi_A}>        \mathcal{M}
\end{CD}
\]
In fibered charts, we have
$$ a = \pi_A (\pi (\hat{X})) = \pi_A (\pi (
\left[ {{\begin{array}{cc}
       X \hfill  \\
       X^5  \hfill  \\
   \end{array} }} \right]))
   = \pi_A (X)
$$
and in the other hand
$$    a = \pi_M (\hat{a}) = 
   \pi_M (\left[ {{\begin{array}{cc}
       a \hfill  \\
       a^4  \hfill  \\
   \end{array} }} \right])
   = \pi_M (\hat{\pi}_A (\hat{X}))
   = \pi_M (\left[ {{\begin{array}{cc}
       \varphi (\hat{X}) \hfill  \\
       \varphi^4 (\hat{X})  \hfill  \\
   \end{array} }} \right])
   = \varphi (\hat{X})
$$
that entails 
\begin{equation}
     \varphi (
\left[ {{\begin{array}{cc}
       X \hfill  \\
       X^5  \hfill  \\
   \end{array} }} \right]) = \pi_A (X) 
\label{varphi(hat(X)) = pi_1 (X)}
\end{equation}
The vector $\hat{\bm{U}}$ tangent to the trajectory of a particle $\hat{\bm{a}}$ is such that 
$$    (T \hat{\pi}_A)  \cdot \hat{\bm{U}} = \bm{0}
$$
In fibered charts, owing to (\ref{varphi(hat(X)) = pi_1 (X)}), this relation is represented by
$$ \hat{M} \, \hat{U} =
    \left[ {{\begin{array}{cc}
       M \hfill &  0 \hfill \\
       \Phi \hfill &  \mu \hfill \\
   \end{array} }} \right] \, 
   \left[ {{\begin{array}{cc}
       U \hfill  \\
       U^5  \hfill  \\
   \end{array} }} \right]
   = \left[ {{\begin{array}{cc}
       0 \hfill  \\
       0  \hfill  \\
   \end{array} }} \right]
$$
where $\Phi$ is a 4-row and $\mu$ is a non null scalar (for $\hat{M}$
being of full rank). In contrast, if $\mu = 0$, a non null solution is $U = 0, \hat{U}^5 \neq 0$, then $\hat{U}$ is not timelike. In the sequel, we exclude this case. 

\vspace{0.3cm}

\textbf{Conformation tensor}

\vspace{0.3cm}

All we have just done is also valid in fibered bases, not necessarily associated to local coordinates. As in hypothesis {\bf (H2)} of Section \ref{SubSection Hypotheses on the pullback connection}, let us consider a fibered basis $(\hat{\bm{e}}'_\alpha) = (\hat{\bm{e}}'_1, \cdots, \hat{\bm{e}}'_4, \bm{\hat{\Omega}}_0)$ of $T_{\hat{\bm{X}}} \hat{\mathcal{U}}$ such that $(\bm{e}'_1, \cdots, \bm{e}'_4)$ is an orthonormal basis and in which the tangent map $T \hat{\pi}_A$ is represented by the matrix
\begin{equation}
     \hat{M}' =
    \left[ {{\begin{array}{cc}
       M' \hfill &  0 \hfill \\
       \Phi' \hfill &  \mu' \hfill \\
   \end{array} }} \right] 
\label{hat(M)' =}
\end{equation}
where $M'$ is full rank and $\mu' \neq 0$. Let us verify now that 7 is the number of independent invariants of $\hat{M}'$ for the action of $\hat{\mathbb{G}}_0$ 
$$ \hat{M}' \mapsto  \hat{M}' \, \hat{P}
$$
We determine the isotropy group of $\hat{M}'$ by solving the system of equations $\hat{M}' = \hat{M}' \, \hat{P}$ with respect to $\hat{P}$ given by (\ref{C & P = for hat(G)_0}), itemized as
$$ M' = M' \, P, \qquad
   \Phi' = \Phi' \, P + \mu' \, b^*, \qquad
   \mu' = \mu'
$$
where $P$ is a Lorentz transformation and the last equation is trivially fulfilled. Like in Section \ref{SubSection - Coadjoint orbit}, we differentiate these equations with respect to $\hat{P}$ at the identity and we take into account that $P$ is skew-adjoint, that leads to determine the non vanishing solutions $(\delta P, \delta b)$ of the linear system
\begin{equation}
     M' \, \delta P = 0 , \qquad 
   \Phi' \, \delta P + \mu' \, \delta b^* = 0
\label{M delta P = 0 & Phi delta P + mu delta b^* = 0}
\end{equation}
where $\delta P^* = -  \, \delta P$. According to Theorem \ref{thm vector product} $\diamondsuit$, if $\delta P \neq 0$ there exist two linear independent vectors $V_1, V_2 \in \mathbb{R}^4$ such that $\delta P = \mathcal{J} (V_1, V_2)$. if $M'^*_1, M'^*_2, M'^*_3$ are the rows of $M'$, then  the first equation is satisfied if $M'^*_k \mathcal{J} (V_1, V_2) = - (\mathcal{J} (V_1, V_2) M_k)^* = 0$ for $1\leq k \leq 3$, that implies there exists a linear combination between the rows of $M'$ and $rank(M') < 3$, in contradiction with the  hypothesis  of full rank, then $\delta P = 0$. As $\mu \neq 0$, the second equation (\ref{M delta P = 0 & Phi delta P + mu delta b^* = 0}) shows that $\delta b = 0$. The isotropy group of $M'$ is reduce to the identity. The dimension of the orbit is
$$ \mbox{dim} \, (\mbox{orb} (\hat{M}')) 
 = \mbox{dim} \,  \hat{\mathbb{G}}_0 - \mbox{dim} (\mbox{iso} (\hat{M}')) = 10 - 0 = 10
$$
The dimension of the set $V_M$ of matrices (\ref{hat(M)' =}) is $3 \times 4 + 4 + 1 = 17$ (the number of non vanishing components).
The number of independent invariants of the orbit is
$$ n_I = \mbox{dim} \, V_M - \mbox{dim} (\mbox{orb} (\hat{M}')) 
       = 17 - 10 = 7
$$
A set of independent invariants are $\mu$ and the 6 independent elements of the $3 \times 3$ matrix
$$ H' = M' \, M'^* =  M' (G'_0)^{- 1} M'^T G_m
$$
in which $G'_0$ is Minkowski's metric (\ref{G'_0 =}) and $G_m$ is Gram's matrix of $\bm{G}_m$.  As in \cite{GR},  we suppose that the behavior of the matter is \textbf{isotropic}
$$ G_m = \mathbb{R}^3, \qquad 
   M^* = G^{-1} M^T, \qquad
   H = M \, M^* = M \, G^{-1} M^T
$$
The matrix  $H$ represents in a local frame of $\mathcal{M}$ a self-adjoint tensor 
$$ \bm{H} = T \pi_A (T \pi_A)^*
$$
called \textbf{conformation} in \cite{GR}.

As in Section \ref{SubSection Explicit form of the connection and the equation of motion}, 
we would like now to work with a coordinate frame 
$(\bm{\partial}_1, \cdots, \bm{\partial}_4)$  
and the corresponding fibered frame
adapted to the section $f$, obtained by the change of basis 
(\ref{change of basis}) with the transformation matrix $\hat{P}$ given by (\ref{hat(P) = ((P , O), (A^* , 1))}) and in which $T \hat{\pi}_A$ is represented by the matrix
$$ \hat{M} = \hat{M}' \hat{P}
    = \left[ {{\begin{array}{cc}
       M \hfill &  0 \hfill \\
       \Phi - 2 \, \mu \, A^* \hfill &  \mu \hfill \\
   \end{array} }} \right] 
$$
where $M = M' P$, $\Phi = \Phi' P$ and $\mu = \mu'$. If $\mu \neq 0$, the matrix $\hat{M}$ being known, the vector $\hat{U}$ tangent to the trajectory is obtained by solving the equation $M \, U = 0$, next by calculating   $\hat{U}^5 = (\Phi - 2 \, \mu \, A^*) \, U / \mu$. Later on, as we are only interested by the trajectory in the space-time when we reduce the problem by pullback as in Section \ref{SubSection Explicit form of the connection and the equation of motion}, we can consider only the conformation as invariant, assuming without loss of generality that $\mu = \partial_5 a^4 = 0$, $\Phi = 2 \, \mu \, A^*$, then the value of $\hat{U}^5$ is arbitrary. 

\vspace{0.3cm}

\textbf{Flux of a physical quantity}

\vspace{0.3cm}

Now we follow the construction proposed in (36.7) of \cite{GR}. Let $\bm{a} \mapsto r_0 (\bm{a})$ be a non-vanishing scalar field on $\mathcal{M}$ representing the density of an extensive physical quantity (for instance, the mass or the electric charge), $vol$ be the Riemannian volume 4-form on the space-time $\mathcal{U}$, $vol_m$ be the Riemannian volume 3-form on the matter manifold $\mathcal{M}$, and $vol_{r_0}$ be the volume 3-form $r_0 \, vol_m$. We define on the space-time the vector field $\bm{J}_r$ such that 
\begin{equation}
   \iota_{\bm{J}_r} vol = \pi^*_A (vol_{r_0})
\label{iota_(vec(J)_r) vol = pi^*_m (r vol_m)}
\end{equation}
In local chart, for any $d_1 X, d_2 X, d_3 X \in \mathbb{R}^4$
\begin{equation}
     vol(J_r, d_1 X, d_2 X, d_3 X)  
  = vol_{r_0} (M \, d_1 X, M \,  d_2 X, M \,  d_3 X) 
\label{vol(J_r, d_1 X, d_2 X, d_3 X- =}
\end{equation}
In Chapter 5, \S 14, Theorem (16) of \cite{CL}, it is proved that this vector exists and is unique. As $r_0 \neq 0$, (\ref{vol(J_r, d_1 X, d_2 X, d_3 X- =}) shows that $\bm{J}_r$ vanishes only if the rank of $M$ is less than 3. Otherwise, in choosing $M d_1 X = J_r$, we see that 
\begin{equation}
 \frac{\partial a}{\partial X} \, J_r 
    = M \, J_r = 0
\label{M J_r = 0}
\end{equation}
then  $\bm{J}_r (\bm{X})$ is tangent to the trajectory passing by $\bm{X}$. In (36.13) of \cite{GR}, it is proved that
\begin{theorem}
If $J_r$ is not lightlike 
\begin{itemize}
    \item[(i)] we have
    $$ (J_r)^* J_r = - (r_0)^2  \det{H}
    $$ 
    then the conformation matrix  $H = M \, M^*$ is regular
    \item[(ii)] the orthogonal projection matrix onto 
    $J_r$ is
    $$ \frac{J_r (J_r)^*}{(J_r)^* J_r} 
       = 1_{\mathbb{R}^4} - M^* H^{-1} M
    $$ 
\end{itemize}
\label{Theorem Conformation & Orthogonal projection}
\end{theorem}
Assuming that $J_r$ is timelike, we deduce from (i) that
$$ h = - \det{H} > 0
$$
Introducing
$$ r = r_0 \sqrt{h}
$$
the unique vector $\bm{J}_r$ satisfying (\ref{iota_(vec(J)_r) vol = pi^*_m (r vol_m)}) is the tangent vector to the trajectory of which the length is given by  $(J_r)^* J_r = r^2$, then we can put $J_r = r \, U$ with $U^* U = 1$, representing the 4-\textbf{flux of} $r$
$$ \bm{J}_r = r \, \bm{U}
    = r \, \frac{d\bm{X}}{ds}  
$$
where $r$ is a short form of $r_0 (\pi_A (\bm{X})) \, \sqrt{\det{(T \pi_A (T \pi_A)^*)}}$.
Reasoning as in the proof of Theorem \ref{thm stationary action principle} and owing to (\ref{iota_(vec(J)_r) vol = pi^*_m (r vol_m)}), we have
$$ \mbox{div} \, (\bm{J}_r) \, vol
  = \mathcal{L}_{\bm{J}_r} \, vol
  =  d \lbrack  \iota_{\bm{J}_r} vol \rbrack
  = \pi^*_A ( d (vol_{r_0}) ) = 0
$$
because $d (vol_{r_0})$ is a 4-form field on $\mathcal{M}$ of dimension 3, then we obtain the \textbf{conservation identity of the} 4-\textbf{flux of} $r$
\begin{equation}
     \mbox{div} \, \bm{J}_r = 0
\label{conservation identity of the 4-flux of r}
\end{equation}

\vspace{0.3cm}

\textbf{Lagrangian of the matter}

\vspace{0.3cm}

On this ground, our aim is to construct a Lagrangian $L_M$ for the matter that is \textbf{invariant} by every diffeomorphism $\varphi$ of the space-time $\mathcal{U}$
$$ L_M (\bm{a}, T \pi_A, \bm{G}, \bm{A}^*) = L_M (\bm{a}, T \pi_A \circ T \varphi, \varphi^* (\bm{G}), \varphi^*(\bm{A}^*)) 
$$
where $\varphi^*$ stands for the pullback by $\varphi$ and $\bm{A}^*$ is the adjoint of $\bm{A}$. In local charts, for every transformation matrix $P$
$$ L_M (a, M, G, A^*) = L_M (a', M', G', A'^*) , \quad
 a = a', \quad 
 M' = M \, P, \quad
 G' = P^T G \, P, \quad
 A'^* = A^* P
$$
Particles without spin are characterized by two physical quantities, their mass (or energy if $c = 1$) and their electric charge. Hence, the Lagrangian must at least depend on the densities of mass and charge. According to the principle of stationary action \textbf{(P2)}, we adopt the decomposition
$$ L_M (a, M, G, A^*)  = \kappa \, \lbrack L_m (a, M, G)  + L_e (a, M, G, A^*) \rbrack
$$
where $\kappa$ is a non vanishing \textbf{coupling constant}, $L_m$ is active for all particles, charged or not, sensitive to the gravitation forces, while $L_e$ is active only for charged particles, sensitive also to the electromagnetic forces. For $L_m$ being invariant, it must depend on $a, M, G$ through the invariants $a$ and $H$. At this stage, we restrict our presentation to fluids then the dependence on $H$ can be limited to a dependence on $h = - \det H$ only. The expression
$$ L_m (a, M, G) = \rho_{m0} (a) \, \lbrack \sqrt{h} + \psi (h) \rbrack
$$
contains the \textbf{mass density}
$$ \rho_m = \rho_{m0} (a) \, \sqrt{h}
$$
that can be completed by an internal energy to modelize pressure effects through the function $\psi$. For $L_e$ being invariant, it  must depend on the invariants constructed from $a, M, G, A^*$. We already know the invariants $a$ and $h$ to which we have to add a joint invariant of $M$ and $A^*$. The \textbf{electric current density} $\bm{J}_{\rho_e}$ is the 4-flux of the \textbf{electric charge density}
\begin{equation}
	 \rho_e = \rho_{e0} (a) \, \sqrt{h}
\label{electric charge density}
\end{equation}
defined by, according to (\ref{vol(J_r, d_1 X, d_2 X, d_3 X- =})
\begin{equation}
	      vol(J_{\rho_e}, d_1 X, d_2 X, d_3 X)  
  = vol_{\rho_{e0}} (M \, d_1 X, M \,  d_2 X, M \,  d_3 X) 
\label{vol(J_(rho_e), d_1 X, d_2, X, d_3 X) =vol_(rho_e0) (M d_1  X, M d_2 X, M d_3 X)}
\end{equation}
On the other hand, 
$$  \rho_{e0} \, \det 
\left[ \begin{array}{c}
       M \hfill  \\
       A^*  \hfill  \\
   \end{array}  \right]
    vol(J_{\rho_e}, d_1 X, d_2 X, d_3 X)  
$$
$$  = - \rho_{e0} \,  vol \left(
  \left[ \begin{array}{c}
       A^* J_{\rho_e} \hfill  \\
       M \, J_{\rho_e}  \hfill  \\
   \end{array}  \right] ,  
  \left[ \begin{array}{c}
       A^* d_1 X, \hfill  \\
       M \, d_1 X,  \hfill  \\
   \end{array}  \right] ,
     \left[ \begin{array}{c}
       A^* d_2 X, \hfill  \\
       M \, d_2 X,  \hfill  \\
   \end{array}  \right] ,  
     \left[ \begin{array}{c}
       A^* d_3 X, \hfill  \\
       M \, d_3 X,  \hfill  \\
   \end{array}  \right] ,    
  \right)
$$
Taking into account the property (\ref{M J_r = 0}), we have $M \, J_{\rho_e} = 0$ then
$$  \rho_{e0} \, \det 
\left[ \begin{array}{c}
       M \hfill  \\
       A^*  \hfill  \\
   \end{array}  \right]
    vol(J_{\rho_e}, d_1 X, d_2 X, d_3 X)  
=  (A^* J_{\rho_e}) \,   vol_{\rho_{e0}} (M \, d_1 X, M \,  d_2 X, M \,  d_3 X) 
$$
Owing to (\ref{vol(J_(rho_e), d_1 X, d_2, X, d_3 X) =vol_(rho_e0) (M d_1  X, M d_2 X, M d_3 X)}), we construct a function
$$  L_e (a, M, G, A^*)  =
 \rho_{e0} (a)\, \det 
\left[ \begin{array}{c}
       M \hfill  \\
       A^*  \hfill  \\
   \end{array}  \right]
   = - A^* J_{\rho_e} 
   = - \rho_e  \, A^* U
$$
which is invariant because $A'^* U' = (A^* P) (P^{- 1} U) = A^* U$ and the electric charge density is invariant.

In a nutshell, the \textbf{Lagrangian of the matter} reads
$$ L_M (a, M, G, A^*)  = \kappa \, \lbrack (\rho_{m0} - \rho_{e0} \, A^* U) \, \sqrt{h} + \rho_{m0}  \, \psi(h)
\rbrack
$$
which can be enriched to modelize more complex behaviors such as, for instance, those of hyperelastic solids for which the Lagrangian depends on the invariant conformation $H$ in a general way, not only through $h$.

\vspace{0.3cm}

\textbf{Principle of stationary action and law of conservation}

\vspace{0.3cm}

Owing to (\ref{delta L_(z) = p delta z + sigma^j delta (partial z) + E^jk_(z) delta _jk + tilde(T)^^k_(z) A_k}),  the components of the vector $\tilde{\bm{T}}_M$ read
\begin{equation}
    (\tilde{T}_M)^i = - \kappa \,  (J_{\rho_e})^i 
                    = - \kappa \, \rho_e U^i
\label{(tilde(T)_M)^k = - chi J_(rho_e)}
\end{equation}
Applying the principle of stationary action, Corollary \ref{Corollary} gives
\begin{equation}
	 \mbox{div} \bm{J}_{\rho_e} = 0
\label{div vec(J)_(rho_e) = vec(0)}
\end{equation}
which is satisfied because of the conservation of the 4-flux of electric charge density $\rho_e$, according to (\ref{conservation identity of the 4-flux of r}).

To obtain the law of conservation, we have to determine the expression of $E^{jk}$ by differentiation the Lagrangian with respect to the metric, {\it i.e.} through $h$, then denoting $\partial_h$ the partial derivative with respect to $h$
$$ \delta L_M = \partial_h L_M \delta h 
    = - \partial_h L_M \delta (\det H)
    = h \, \partial_h L_M Tr (H^{- 1} \delta H)
$$
$$ \delta L_M  
   = h \, \partial_h L_M Tr (H^{- 1} M \delta (G^{-1}) M^T)
 = - h \, \partial_h L_M Tr (M^T H^{- 1} M G^{-1} (\delta G) G^{-1} )
$$
Using Theorem \ref{Theorem Conformation & Orthogonal projection}) (ii) with $J_{\rho_m} = \rho_m U$, it holds
$$ \delta L_M 
 = - h \, \partial_h L_M Tr (G^{-1} (1_{\mathbb{R}^4} - U \, U^*) G^{-1} \delta G)
 =  h \, \partial_h L_M Tr ( (U\, U^T - G^{-1}) \delta G)
$$
or in tensor notation
$$ E^{jk} = h \, \partial_h L_M (G^{jk} - U^j U^k)
$$
from which we deduce the components (\ref{T^jk_(z) = 2 E^jk_(z)  + L_(z) G^jk}) of the tensor field $\bm{T}_M$
\begin{equation}
 (T_M)^{ij} =  2 \, E^{ij}_M + L_M G^{ij}
                   = \kappa \, \lbrack   (\rho + p) U^i U^j - p \, G^{ij} \rbrack
\label{(T_M)^(jk) =}
\end{equation}
in which occur the \textbf{density of energy}
$$ \rho = L_M / \kappa = (\rho_{m0} - \rho_{e0} \, A^* U) \, \sqrt{h} + \rho_{m0}  \, \psi(h)
             = \rho_m + \rho_{m0}  \, \psi(h) - A^* J_{\rho_e}
$$
where the second and third terms represent respectively the internal energy and the coupling with the electromagnetism,
and the \textbf{pressure}
$$ p = (2 \partial_h L_M  - L_M) / \kappa = \rho_{m0} \, (2 \, h \, \partial_h \psi - \psi)
$$
The law of conservation (\ref{law of conservation}) for the matter 
$$ \nabla_i (T_M)^i_j + (\tilde{T}_M)^k F_{kj} = 0
$$
itemizes, owing to (\ref{(tilde(T)_M)^k = - chi J_(rho_e)}) and (\ref{(T_M)^(jk) =}), into
\begin{equation}
	 \nabla_i \lbrack (\rho + p) U^i U_j - p \, \delta^i_j \rbrack
     - \rho_e U^k F_{kj} = 0
\label{nabla_i ((rho + p) U^i U_j - p delta^i_j) - rho_e U^k F_(kj) = 0}
\end{equation}

\vspace{0.3cm}

\textbf{Interpretation of the law of conservation in terms of the} $\hat{\mathbb{G}}_0$-\textbf{connection}

\vspace{0.3cm}

Let a 1-contravariant and 1-covariant free divergence tensor field $\hat{\mathcal{\bm{T}}}$ on the space-time $\mathcal{U}$, section of the pullback bundle $f^* (T\hat{\mathcal{U}} \otimes T^*\hat{\mathcal{U}})$ 
$$ \hat{\nabla}_i \hat{\mathcal{T}}^i_\beta
  = \partial _i \hat{\mathcal{T}}^i_\beta
    + \hat{\Gamma}^i_{i\mu} \hat{\mathcal{T}}^\mu_\beta 
    -  \hat{\mathcal{T}}^i_\mu \hat{\Gamma}^\mu_{i\beta} = 0
$$
with the convention that Greek indices run from 1 to 5 and Latin ones run from 1 to 4. For the $\hat{\mathbb{G}}_0$-connection (\ref{hat(Gamma) = ((Gamma 0 (Gamma^5 0))}), $\hat{\Gamma}^\mu_{i5} = 0$ then the previous equation itemizes into
$$ \hat{\nabla}_i \hat{\mathcal{T}}^i_j
  = \partial _i \hat{\mathcal{T}}^i_j
    + \Gamma^i_{im} \hat{\mathcal{T}}^m_j 
    -  \hat{\mathcal{T}}^i_m \Gamma^m_{ij} 
    -  \hat{\mathcal{T}}^i_5 \Gamma^5_{ij} 
    = \nabla_i \hat{\mathcal{T}}^i_j 
    - \hat{\mathcal{T}}^i_5 \Gamma^5_{ij}= 0
$$
\begin{equation}
	 \hat{\nabla}_i \hat{\mathcal{T}}^i_5
  = \partial _i \hat{\mathcal{T}}^i_5
    + \Gamma^i_{im} \hat{\mathcal{T}}^m_5 
    = \nabla_i \hat{\mathcal{T}}^i_5 = 0
\label{hat(nabla)_i hat(T)^i_j = 0 & hat(nabla)_i hat(T)^i_5 = 0}
\end{equation}
where $\nabla$ is Levi-Civita connection. It is worth to point out that the components $\mathcal{T}^5_\beta$ are absent of these expressions. Mimicking the linear 5-momentum of an elementary particule defined by (\ref{hat(Pi) = ( Pi^* , q)}) and (\ref{Pi^* = m_0 U^* + q A^*}), we claim that the components of the \textbf{linear} 5-\textbf{momentum of a continuum} are
$$ \hat{\Pi}_j = \rho U_j - 2 \, \rho_e A_j, \qquad
     \hat{\Pi}_5 = \rho_e
$$
Adding a projection term to take into account the effects of the pressure $p$, we claim that
$$ \hat{\mathcal{T}}^i_j  = U^i \hat{\Pi}_j + p \, (U^i U_j - \delta^i_j)
     =   U^i (\rho U_j - 2 \, \rho_e A_j) + p \, (U^i U_j - \delta^i_j) , \qquad
     \hat{\mathcal{T}}^i_5 = U^i \hat{\Pi}_5 = \rho_e U^i
$$
Owing to (\ref{Gamma^5_(ij) = F_(ij) - 2 nabla_i A_j}), the equation
$$ \mbox{div} \, \hat{\mathcal{\bm{T}}} = \bm{0}
$$
is represented in local charts by (\ref{hat(nabla)_i hat(T)^i_j = 0 & hat(nabla)_i hat(T)^i_5 = 0})
$$ \nabla_i \lbrack (\rho + p) U^i U_j - p \, \delta^i_j \rbrack
     - \rho_e U^i F_{ij} = 0, \qquad
     \nabla_i (\rho_e U^i) = 0
$$
then, by comparison to (\ref{div vec(J)_(rho_e) = vec(0)}) and (\ref{nabla_i ((rho + p) U^i U_j - p delta^i_j) - rho_e U^k F_(kj) = 0}), is equivalent to the equations provided by the principle of stationary action and the law of conservation
$$ \mbox{div} \, \tilde{\bm{T}}_M = 0, \qquad 
    \mbox{div} \,  \bar{\bm{T}}_M + \tilde{\bm{T}}_M\cdot \bm{F} = \bm{0}
$$

\subsection{Field equations of the gravitation and electromagnetism}

In general relativity, the gravitation is represented by the geometry of the space-time through the connection $\Gamma$ generated by 10 potentials, the independent components of the metric $G$. In our unified theory of gravitation and electromagnetism, the geometry of the space-time stems from the $\hat{\mathbb{G}}_0$-connection $\hat{\Gamma}$ preserving the pullback $\hat{\bm{G}}_0$ of the space-time metric and the vector $\hat{\bm{\Omega}}_0$. It is generated by 14 potentials, the independent components of $G$ and $A^*$.  

\vspace{0.3cm}

\textbf{Curvature tensor}

\vspace{0.3cm}

Our starting point is to calculate the curvature tensor $\hat{\bm{R}}$ on the space-time $\mathcal{U}$ endowed with the pullback connection. This tensor is defined, for any smooth section $\hat{\bm{W}}$ of $T\hat{\mathcal{U}}$ and for all smooth fields $\bm{X} \mapsto \bm{U}, \bm{X} \mapsto \bm{V}$  of tangent vectors to the space-time, by the relation 
    \begin{equation}
     \hat{\nabla}_{(Tf) \, \bm{U}} \, (\hat{\nabla}_{(Tf) \, \bm{V}} \, \hat{\bm{W}})
   - \hat{\nabla}_{(Tf) \, \bm{V}} \, (\hat{\nabla}_{(Tf) \, \bm{U}}
      \, \hat{\bm{W}})
   - \hat{\nabla}_{(Tf) \,\lbrack \bm{U}, \bm{V}\rbrack} \, \hat{\bm{W}}
   = \hat{\bm{R}} (\bm{U}, \bm{V}) \, \hat{\bm{W}}
    \label{curvature tensor}
    \end{equation}
Then it is a 1-contravariant and 3-covariant tensor, skew-symmetric with respect to the arguments  $\bm{U}$ and $\bm{V}$. With the convention that Greek indices run from 1 to 5 and Latin ones run from 1 to 4, it is represented by the components 
$$ \hat{R}^\alpha_{ij\beta} 
 =  \hat{\Gamma}^\alpha_{i\mu} \hat{\Gamma}^\mu_{j\beta}
  - \hat{\Gamma}^\alpha_{j\mu} \hat{\Gamma}^\mu_{i\beta}
  + \partial_i \hat{\Gamma}^\alpha_{j\beta}
  - \partial_j \hat{\Gamma}^\alpha_{i\beta}
$$
skew-symmetric with respect to the indices $i$ and $j$. Owing to the form (\ref{hat(Gamma) (U) =}) of the connection matrix, Christoffel's symbols with a covariant index equal to 5 vanish  then the only non null curvature components are
$$ R^p_{ijk} = \hat{R}^p_{ijk} 
 =  \Gamma^p_{i\mu} \Gamma^\mu_{jk}
  - \Gamma^p_{j\mu} \Gamma^\mu_{ik}
  + \partial_i \Gamma^p_{jk}
  - \partial_j \Gamma^p_{ik}
$$
$$ \tilde{R}_{ijk} = \hat{R}^5_{ijk} 
 =  \Gamma^5_{i\mu} \Gamma^\mu_{jk}
  - \Gamma^5_{j\mu} \Gamma^\mu_{ik}
  + \partial_i \Gamma^5_{jk}
  - \partial_j \Gamma^5_{ik}
$$
The components $R^p_{ijk}$ are the ones of the classical curvature tensor of space-time endowed with the metric $\bm{G}$ (hypothesis {\bf H1}). Developing (\ref{Gamma^5_(ij) = F_(ij) - 2 nabla_i A_j}), one has
$$ \Gamma^5_{ij} = -  \partial_i A_j  
  -  \partial_j A_i
  + 2 \,  \Gamma^k_{ij} A_k
$$
The additional curvature tensor components read
\begin{equation}
     \tilde{R}_{ijk} 
 =  \nabla_k F_{ji} + 2 A_q R^q_{ijk}
\label{tilde(R)_(ijk) = nabla_k F_(ij) +  2 A_q R^q_(ijk)}
\end{equation}

\vspace{0.3cm}

\textbf{Lagrangian of the geometry}

\vspace{0.3cm}

According to  Chapter 5, \S 14, (35.8) to (35.10) of \cite{GR}, the Lagrangian of the geometry depends on the connection and its partial derivatives through the curvature tensor
$$ L_G (\hat{\Gamma}, \partial_i \hat{\Gamma}, \bm{G}, \bm{A}^*)
    = L_G (\hat{\Gamma}, \hat{\bm{R}}, \bm{G}, \bm{A}^*)
$$
We assume the decomposition 
$$ L_G (\hat{\Gamma}, \hat{\bm{R}}, \bm{G}, \bm{A}^*) =
    L_g (\Gamma, \bm{R}, \bm{G}) 
    + L_{em} (\Gamma^5, \tilde{\bm{R}}, \bm{G}, \bm{A}^*)
$$
where $L_g$ describes the behavior of the gravitation alone and $L_{em}$  is an extra term to take into account the one of the electromagnetism. The Lagrangian must be invariant for every diffeomorphism of the space-time. 
Introducing the Ricci tensor of components obtained by contraction 
$$ R_{jk} = R^p_{pjk}
$$
the most simple invariant built from $R^p_{ijk}$ and $G^{ij}$ is the scalar curvature 
$$ R = G^{jk} R_{jk}
$$
The Hilbert-Einstein Lagrangian including the \textbf{cosmological constant} $\Lambda$ is 
$$ L_g (\Gamma, \bm{R}, \bm{G}) 
    = - \Lambda + \frac{1}{2} \, G^{ij} R_{ij}
$$
Besides, from $\tilde{R}_{ijk}, G^{ij}$ and $A_i$, we can construct the invariant
\begin{equation}
     \tilde{R} = A_r G^{ri} G^{jk} \tilde{R}_{ijk} = A_r \tilde{R}^{rk}_{\;\;\;\,k}
\label{invariant tilde(R) =}
\end{equation}
Then we consider the Lagrangian of the electromagnetism
$$ L_{em} (\Gamma^5, \tilde{\bm{R}}, \bm{G}, \bm{A}^*)
    = - \tilde{k}\, A_r G^{ri} G^{jk} \tilde{R}_{ijk} 
$$
where $\tilde{k}$ is a new coupling constant.

In a nutshell, the Lagrangian of the geometry reads
$$  L_G (\hat{\Gamma}, \hat{\bm{R}}, \bm{G}, \bm{A}^*) 
    = - \Lambda + \frac{1}{2} \, G^{ij} R_{ij}
    - \tilde{k}\, A_r G^{ri} G^{jk} \tilde{R}_{ijk}
$$
from which we deduce the components (\ref{T^jk_(z) = 2 E^jk_(z)  + L_(z) G^jk}) of the tensor field $\bm{T}_G$
\begin{equation}
 (T_G)^{ij} =  2 \, (E_G)^{ij} + L_G G^{ij}
    = - \lbrack R^{ij} - \frac{1}{2} \, R \, G^{ij} + \Lambda \, G^{ij}
        \rbrack 
     + \tilde{k} \, \lbrack A^{( i} \, \tilde{R}^{j)\;\,k}_{\;\;\;k}  
                           + A_r \tilde{R}^{r(ij)} 
    - \frac{1}{2} \, \tilde{R} \, G^{ij}        \rbrack 
\label{(T_G)^(jk) =}
\end{equation}
Besides, the components of the vector $\tilde{\bm{T}}_G$ are
\begin{equation}
          (\tilde{T}_G)^i = - \tilde{k} \, \left( \nabla_j F^{ji} + 2 A_q R^{qij}_{\quad\;j} \right)
\label{(tilde(T)_G)^i =}
\end{equation}

\vspace{0.3cm}

\textbf{Field equations}

\vspace{0.3cm}

Applying the stationary action principle (Theorem \ref{thm stationary action principle}, (ii) and (iii)), we have
$$ \bm{T} =  \bm{T}_G + \bm{T}_M = \bm{0}, \qquad
    \tilde{\bm{T}} = 
    \tilde{\bm{T}}_G + \tilde{\bm{T}}_M = \bm{0}
$$
which, owing to (\ref{(tilde(T)_M)^k = - chi J_(rho_e)}), (\ref{(T_M)^(jk) =}), (\ref{(T_G)^(jk) =}) and (\ref{(tilde(T)_G)^i =}), leads to
\begin{equation}
      R^{ij} - \frac{1}{2} \, R \, G^{ij} + \Lambda \, G^{ij} 
    - \tilde{k} \, \lbrack A^{( i} \, \tilde{R}^{j)\;\,k}_{\;\;\;k}  
                           + A_r \tilde{R}^{r(ij)} 
    - \frac{1}{2} \, \tilde{R} \, G^{ij}        \rbrack 
= \kappa \, \lbrack   (\rho + p) U^i U^j - p \, G^{ij} \rbrack 
\label{field equation derived from G_(ij)}
\end{equation}
\begin{equation}
      - \tilde{k} \, \lbrack \nabla_j F^{ji} + 2 A_q R^{qij}_{\quad j} \rbrack
    = \kappa \, \rho_e U^i  
\label{epsilon_0 (nabla_j F^(ji) + 2 A_q R^(qij)_j + rho_e U_i = 0}
\end{equation}

In the absence of the electromagnetic potential, the first equation is reduced to
$$    R^{ij} - \frac{1}{2} \, R \, G^{ij} + \Lambda \, G^{ij} 
= \kappa \, \lbrack   (\rho + p) U^i U^j - p \, G^{ij} \rbrack 
$$
We recover Einstein's equations for the gravitation. Einstein's constant (where $c = 1$  and the gravitational constant $G_N$ with the index $N$ in Newton's honour) 
$$ \kappa = 8 \, \pi \, G_N
$$
has been chosen such that the non-relativistic limit yields the usual form of Newton's law of universal gravitation.

Our aim now is to discuss the {\bf classical limits} of these field equations, starting with (\ref{epsilon_0 (nabla_j F^(ji) + 2 A_q R^(qij)_j + rho_e U_i = 0}). First, we consider the former term. In the Galilean approximation (\cite{Duval 1985, Duval 1991, Souriau 1997a, AffineMechBook}), 
the connection is associated to Galilei group and Christoffel's symbols $\Gamma^m_{ij}$ vanish except\footnote{with the convention that the index $t$ corresponds to time and  $a, b, c, d$ to the spatial coordinates.}
\begin{equation}
   \Gamma^a_{tt} = - g^a,\qquad 
   \Gamma^a_{tb} = \Gamma^a_{bt} = \Omega^a_b\ ,
\label{Gamma^j_00 = - g^j & Gamma^j_0k = Omega^j_k} 
\end{equation}
where occur the gravity $g^a$ and Coriolis' effects given by  $\Omega^a_b$, a simplified notation for the elements of the $3 \times 3$ skew-symmetric matrix $j (\Omega)$ such that $j (\Omega) \, v = \Omega \times v$ for all 3-columns $\Omega$ and $v$. For a 2-contravariant tensor of components $F^{ji}$, the covariant divergence reads
$$ \nabla_j F^{ji} =
   \partial_j F^{ji} 
   + \Gamma^j_{jm} F^{mi}
   + F^{jm} \Gamma^i_{jm} 
$$
that, owing to (\ref{Gamma^j_00 = - g^j & Gamma^j_0k = Omega^j_k}), itemizes respectively for $j = t$ and $j = a$ as
$$ \nabla_j F^{jt} =
   \partial_j F^{jt} , \qquad
   \nabla_j F^{ja} =
   \partial_j F^{ja} 
   - g^a F^{tt}
   + (F^{bt} + F^{tb}) \Omega^a_b 
$$
For the electromagnetic field, $F^{ij} = - F^{ji}$, this leads, at the Galilean approximation, to the exact simplification
\begin{equation}
     \nabla_j F^{ji} 
   = \partial_j F^{ji} 
\label{nabla_j F^(ji) = partial_j F^(ji)}
\end{equation}

Next, we would like to assess the order of magnitude of the second term with respect to the first one in (\ref{epsilon_0 (nabla_j F^(ji) + 2 A_q R^(qij)_j + rho_e U_i = 0}), {\it i.e.} the coupling term between gravitation and electromagnetism containing 
$$ A_q R^{qij}_{\quad j} = 
    A_q G^{im} G^{jq} R^q_{\;\; mkj}
$$
As we are interested by an estimate, we consider only the gravity. 
At the Newtonian approximation, the only non vanishing Christoffel's symbol is  $\Gamma^a_{tt} = - g^a$ and 
$$  G^{- 1} \cong G \cong diag (1, -1,  -1, -1)
$$
then
$$ A_q R^{qij}_{\quad j} \cong 
    - A_a G^{ib} \partial_b g^a
$$
which, introducing the gravity potential $\varphi$, itemizes as 
\begin{equation}
     A_q R^{qtj}_{\quad j} \cong 0, \qquad
   A_q R^{qbj}_{\quad j} \cong 
     A_a \delta^{bc} \delta^{ad}  \partial_c \partial_d \varphi
\label{A_q R^(qtj)_j cong 0 &}
\end{equation}
For $i = t$ in (\ref{epsilon_0 (nabla_j F^(ji) + 2 A_q R^(qij)_j + rho_e U_i = 0}), the equation is reduced, at the Newtonian approximation, to 
$$ - \tilde{k} \,  \partial_j F^{jt}  
    = \kappa \, \rho_e 
$$
or
$$ \tilde{k} \, \mbox{div} \, E = \kappa \, \rho_e
$$
Provided
\begin{equation}
     \tilde{k} = \kappa \, \epsilon_0 = 8 \, \pi \, G_N \, \epsilon_0
\label{tilde(k) = kappa epsilon_0 = 8 pi G_N epsilon_0}
\end{equation}
where $\epsilon_0$ is the \textbf{permittivity} occurring in Coulomb's law of electrostatic, it is just Maxwell-Gauss equation
$$ \mbox{div} \, E = \frac{\rho_e}{\epsilon_0}
$$
Let $\bar{L}, \bar{A}, \bar{\varphi}$ the characteristic length, electromagnetic potential and gravity potential. Owing to (\ref{nabla_j F^(ji) = partial_j F^(ji)}) the first term of (\ref{epsilon_0 (nabla_j F^(ji) + 2 A_q R^(qij)_j + rho_e U_i = 0}) is of the order of $\bar{L}^{- 2} \bar{A}$ while, owing to (\ref{A_q R^(qtj)_j cong 0 &}), the second term is of the order of $\bar{L}^{- 2} \bar{A} \, \bar{\varphi}$, then can be neglected because, at the Newtonian approximation, $\bar{\varphi} \ll 1$ (with units such that $c = 1$) and (\ref{epsilon_0 (nabla_j F^(ji) + 2 A_q R^(qij)_j + rho_e U_i = 0}) is reduced to the second group of {\bf Maxwell equations} (Maxwell-Gauss and Maxwell-Amp\`{e}re laws)
$$ \mbox{div} \, E = \frac{\rho_e}{\epsilon_0}, \qquad \mbox{curl} \, B = \partial_t E  + \mu_0 \rho_e v^i
$$
where $\mu_0 = (\epsilon_0)^{-1}$ is the \textbf{permeability} (bearing in mind that $c = 1$),
while, because the 2-form $\bm{F}$ is closed, its exterior derivative is zero and we recover the first  group (Maxwell-Faraday and Maxwell-Thomson laws) 
$$ \mbox{curl} \, E +   \partial_t B  = 0, \qquad \mbox{div} \, B = 0
$$

\textbf{Remark 1.} The previous approximation is rough. In many situation, it can be improved, depending on the reference length,  notably when the gravity can be considered on the Earth as constant.

\textbf{Remark 2.} Besides, incorporating Coriolis' effects in the analysis of orders of magnitude leads to more complex expressions but the conclusion concerning the order of the second term of (\ref{epsilon_0 (nabla_j F^(ji) + 2 A_q R^(qij)_j + rho_e U_i = 0}) remains. 

\vspace{0.3cm}

Next, we look back to the \textbf{classical limit of the former field equation} (\ref{field equation derived from G_(ij)}), taking into account (\ref{tilde(k) = kappa epsilon_0 = 8 pi G_N epsilon_0})
$$    R^{ij} - \frac{1}{2} \, R \, G^{ij} + \Lambda \, G^{ij} 
    - 8 \, \pi \, G_N \epsilon_0 \, \lbrack A^{( i} \, \tilde{R}^{j)\;\,k}_{\;\;\;k}  
                           + A_r \tilde{R}^{r(ij)} 
    - \frac{1}{2} \, \tilde{R} \, G^{ij}        \rbrack 
= 8 \, \pi \, G_N  \lbrack   (\rho + p) U^i U^j - p \, G^{ij} \rbrack 
$$
To be consistent, the second term in (\ref{tilde(R)_(ijk) = nabla_k F_(ij) +  2 A_q R^q_(ijk)}) is neglected and, at the Newtonian approximation, the invariant (\ref{invariant tilde(R) =}) is reduced to 
$$ \tilde{R} = A_i \partial_j F^{ji}
$$
Being interested by the order of magnitude, we withhold only the term with $A_t = \phi$ in $\tilde{R}$, then
$$ G_N \epsilon_0 \tilde{R} \thickapprox G_N \phi \, \epsilon_0 \, \mbox{div} \, E
    \thickapprox G_N \rho_e \phi
$$
Comparing to the order of magnitude $G_N \rho$ of the right hand member of (\ref{field equation derived from G_(ij)}), we conclude that the coupling term can be neglected. Indeed, $\rho_e \phi$ is the stored electric energy density, very negligible with respect to the energy density $\rho$ (that is $\rho \, c^2$ in SI units).

\textbf{Remark.} Arguably, this particularly very weak coupling term in (\ref{field equation derived from G_(ij)}) can be related to Dirac's Large Number Hypothesis (LNH) that the ratio of the electrical to the gravitational forces between a proton and an electron of electric charge $e$
$$ \frac{F_e}{F_g} = \frac{e^2}{4 \, \pi \, \epsilon_0 G_N m_e m_p}
$$
constitutes a very large dimensionless number, some 40 orders of magnitude today, indeed the coupling constant is 
$$ \tilde{k} = \frac{2 \, e^2}{m_e m_p} \, \frac{F_g}{F_e}
$$
where the inverse ratio $F_g / F_e$ is about $10^{-40}$. Observing the coincidence of the ratio of this force scales and that of the size scales in the Universe, Paul Dirac \cite{Dirac 1937} made the conjecture that physical constants are actually not constant and the gravitational constant is inversely proportional to the age of the Universe. 
The author would also to draw the attention on the coincidence with the square of another dimensionless number, the ratio of the double of the geometric mean of Compton wavelengths of proton and neutron to the size $l_K$ of the space along the fifth dimension 
$$ \frac{F_e}{F_g} = \left( \frac{2 \, \sqrt{\lambda_C \lambda_{C, p}}}{l_K} \right)^2
$$
where, according to  Chapter 7, \S 42, p. 412 of \cite{GR}, in SI units
$$ l_K = 8 \, \pi^{3/2} \, \frac{h \, \sqrt{G_N \epsilon_0}}{e \, c} = 2.38 \times 10^{-31} \, \mbox{cm}
$$
More in-depth discussion should be given in relation with the cosmological scenario for the evolution of elementary particle structure of Section \ref{Section - A cosmological scenario for the evolution of elementary particle structure}.

\vspace{0.3cm}

In contrast, if the classical approximations above are not relevant, the gravitation and electromagnetism fields are coupled through the equations
$$    R^{ij} - \frac{1}{2} \, R \, G^{ij} + \Lambda \, G^{ij} 
    - 8 \, \pi \, G_N \epsilon_0 \, \lbrack A^{( i} \, \tilde{R}^{j)\;\,k}_{\;\;\;k}  
                           + A_r \tilde{R}^{r(ij)} 
    - \frac{1}{2} \, \tilde{R} \, G^{ij}        \rbrack 
= 8 \, \pi \, G_N  \lbrack   (\rho + p) U^i U^j - p \, G^{ij} \rbrack 
$$
$$       \epsilon_0 \, \lbrack \nabla_j F^{ji}
         + 2 A_q R^{qij}_{\quad j} \rbrack
         + \rho_e U^i   = 0 
$$
In particular, in presence of a strong gravitation field, Maxwell equations have to be modified, by considering the covariant derivative instead of the partial derivative.

 \section{Conclusion}

Firstly, we proposed a symmetry group $\hat{\mathbb{G}}_0$ for which the electric charge of a particle is an invariant of its coadjoint orbit, then of which the value does not depend on the observer. Secondly, we constructed a $\hat{\mathbb{G}}_0$-connection allowing to recover the expected equation of motion today of a particle,  including Lorentz force and  proving the charge is an integral of the motion. Finally, we deduced from a 5D extension of the  variational relativity the field equations where, at the classical limit, the coupling term is negligible and we recover Einstein equations for the gravitation and Maxwell equations, without need to introduce a dilaton or to violate an equation as in the classical Kaluza-Klein theory.    

Author's opinion is that these results are strong arguments to claim that  \textbf{the Lie group} $\hat{\mathbb{G}}_0$ \textbf{is the symmetry group of the electrodynamics compatible with the observations today}. 

In contrast, we affirm that the symmetry group $\hat{\mathbb{G}}_1$ of the Kaluza-Klein theory leads to a classification of the elementary particles in the framework of an \textbf{unified theory merging the gravitational and electromagnetic forces relevant to describe the early Universe}. 

\vspace{0.2cm}

The sound idea of a fifth dimension, if used in a suitable geometric formalism taking into account its overwhelmingly small size, seems now more than ever to offer a promising future for new developments among them we can emphasize:

\begin{itemize}
    \item[1.] 
    For the characterization of charged elementary particles, we largely relied on \cite{SSD, SSDEng} without releasing the full potential of the theoretical framework, in particular the geometric quantization. These issues have not been considered yet in this work but would deserve to be investigated.
    \item[2.] 
    The next step would be to extend Kaluza-Klein theory to a non-abelian gauge group as in \cite{Kerner 1968} with the mathematical tools developed in this paper, in particular the coadjoint orbit method.
    \item[3.]
    Going back in time, it would worth to revisit works on the early universe cosmology that we think they did not receive the welcome they deserved because based on Kaluza-Klein theory whose weaknesses were known. They could provide a better understanding for Dirac's large number hypothesis as suggested in \cite{Chodos 1980}  and to offer a resolution to the horizon problem as claimed in \cite{Sahdev 1984}.   
\end{itemize}

\vspace{0.3cm}

\textbf{Acknowledgements}

\vspace{0.3cm}

The author would like to thank Leonid Ryvkin for discussions that helped to clarify certain points and to Richard Kerner whose advice was very valuable to me. He is also grateful to Claude Vall\'ee for drawing author's attention on the reference \cite{Cartan 1934}.

\end{document}